\documentclass[11pt,a4paper]{article}
\usepackage[utf8]{inputenc}
\usepackage[left=1in,right=1in,top=1in,bottom=1in]{geometry}
\usepackage{amsmath}
\usepackage{amsthm}
\usepackage{amsfonts}
\usepackage{amssymb}
\usepackage{scrextend}
\usepackage{mathtools}
\usepackage{algorithm}
\usepackage[noend]{algpseudocode}
\usepackage{float}
\usepackage{multicol}
\usepackage{graphicx}
\usepackage{verbatim}
\usepackage{ctable}
\usepackage{authblk}
\usepackage{makecell}
\usepackage{fancyhdr}
\usepackage{color}
\usepackage{hyperref}
\usepackage[section]{placeins}
\usepackage{lineno}

\begin{document}
\title{Quantifying Dissipation in Actomyosin Networks}
\author[1]{Carlos Floyd}
\author[1,2,3,$\dag$]{Garegin A. Papoian}
\author[1,2,3,4,*]{Christopher Jarzynski}
\affil[1]{Biophysics Program, University of Maryland, College Park, MD 20742 USA}
\affil[2]{Department of Chemistry and Biochemistry, University of Maryland, College Park, MD 20742 USA}
\affil[3]{Institute for Physical Science and Technology, University of Maryland, College Park, MD 20742 USA}

\affil[4]{Department of Physics, University of Maryland, College Park, MD 20742 USA}
\affil[$\dag$]{email: gpapoian@umd.edu}
\affil[*]{email: cjarzyns@umd.edu}

\date{\today}

\begin{titlepage}
\maketitle
\abstract{Quantifying entropy production in various active matter phases will open new avenues for probing self-organization principles in these far-from-equilibrium systems.  It has been hypothesized that the dissipation of free energy by active matter systems may be optimized to produce highly dissipative dynamical states, hence, leading to spontaneous emergence of more ordered states. This interesting idea has not been widely tested. In particular, it is not clear whether emergent states of actomyosin networks, which represent a salient example of biological active matter, self-organize following the principle of dissipation optimization. In order to start addressing this question using detailed computational modeling, we rely on the MEDYAN simulation platform, which allows simulating active matter networks from fundamental molecular principles. We have extended the capabilities of MEDYAN to allow quantification of the rates of dissipation resulting from chemical reactions and relaxation of mechanical stresses during simulation trajectories. This is done by computing precise changes in Gibbs free energy accompanying chemical reactions using a novel formula, and through detailed calculations of instantaneous values of the system's mechanical energy. We validate our approach with a mean-field model that estimates the rates of dissipation from filament treadmilling. Applying this methodology to the self-organization of small disordered actomyosin networks, we find that compact and highly cross-linked networks tend to allow more efficient transduction of chemical free energy into mechanical energy. In these simple systems, we do not observe that spontaneous network reorganizations lead to increases in the total dissipation rate as predicted by the dissipation-driven adaptation hypothesis mentioned above. However, whether such a principle operates in more general, more complex cytoskeletal networks remains to be investigated.

}
\end{titlepage}

\section{Introduction}

The actin-based cytoskeleton is a dynamic supramolecular structure that, by sustaining and releasing mechanical stress in response to various physiological cues, mediates the exertion of force by cells both on their environments and within their bodies \cite{fletcher2010cell, delon2007integrins}.  These cytoskeletal structures are composed of long (on the order of 1 $\mu m$ \textit{in vivo} \cite{burlacu1992distribution}) actin polymers which are interconnected by various cross-linkers, as well as by myosin motor filaments, resulting in a three-dimensional network-like organization referred to as an ``actomyosin network" \cite{fritzsche2016actin, stricker2010mechanics}.\footnote{In our terminology, we will distinguish between ``cross-linking proteins," which will include both active (e.g. myosin filaments) and passive (e.g. $\alpha$-actinin and fascin) proteins that bind to adjacent actin filaments, and ``cross-linkers," which refer exclusively to passive cross-linking proteins.} 
Part of the intricacy of actomyosin network dynamics is due to the mechanosensitive kinetic reaction rates controlling cross-linker and myosin filament unbinding as well as myosin filament walking: at high tension, cross-linkers will unbind more quickly (slip-bond) whereas motors will unbind and walk less quickly (catch-bond and stalling). \cite{wachsstock1993affinity, keller2000mechanochemistry, albert2014stochastic}.  These reactions control the actomyosin network connectivity, which in turn determines the ability of the network to globally distribute stress \cite{alvarado2017force}.  Thus the mechanosensitive feedback introduces nonlinear coupling between the stress sustained by an actomyosin network and the network's ability to reorganize in response to that stress.  In order to be responsive to physiological cues, the dynamics of these systems occur far from thermodynamic equilibrium; the hydrolysis of an out-of-equilibrium concentration of ATP molecules fuels a) the stress-generating activity of the myosin motor filaments, and b) filament treadmilling \cite{kovacs2003functional, zhang2016thermodynamic, erdmann2016sensitivity, mccullagh2014unraveling, li2009actin}.  Filament treadmilling is a steady-state situation in which the polymerization at the plus end of the filament is compensated by the depolymerization at the minus end, resulting in the filament moving forward without its length changing.  As a result of these these local free energy-consuming processes, actomyosin networks constitute an interesting and biologically important example of soft active matter. Active matter is composed of agents that individually transduce free energy from some external source, in this case the chemical potential energy of many ATP molecules \cite{ramaswamy2010mechanics, marchetti2013hydrodynamics, hill2004free}.  Dissipation in these systems results when the free energy consumed $\Delta G$ is greater than the quantity of work $W$ done by the system on its environment, with the remainder $\Delta G - W$ serving to increase the total entropy.

The viewpoint of actomyosin networks as active matter systems has been fruitfully adopted in recent theoretical and experimental studies, yet a lack of ability to quantify the rates of free energy transduction by these systems has hindered development of some of these lines of study.  The emergence of distinctive dynamical states (for instance pulsing actin waves or vortices) during the self-organization of actomyosin systems has been documented in several \textit{in vitro} experiments \cite{tan2016self, kohler2011structure, schaller2010polar}.  These emergent patterns depend sensitively on the concentrations of myosin filaments and cross-linkers:  myosin filament concentration controls the level of active stress generation, and cross-linker concentration controls the degree of mechanical coupling of actin filaments, which has been described using the language of percolation theory \cite{alvarado2013molecular,alvarado2017force}.  While these emergent dynamic patterns have been characterized in detail, a general mechanism explaining why these patterns emerge under given conditions has not yet been proven.  It might be expected, given that these systems operate away from thermodynamic equilibrium, that the quantity of free energy dissipated during a system's evolution is optimized, similar to the principle of minimum entropy production in the near-equilibrium theory of irreversible thermodynamics \cite{prigogine1967introduction}.  However, this minimum entropy production principle breaks down in the far-from-equilibrium, nonlinear-response regime, where many active matter systems including actomyosin networks operate \cite{prigogine1971biological}.  It has recently been proposed that another optimization principle applies arbitrarily far from equilibrium.  This idea, referred to as dissipation-driven adaptation, suggests that, in general, a coarse-grained trajectory of some nonequilibrium system will be more likely than all alternative trajectories if the amount of free energy absorbed and dissipated along that trajectory is maximal \cite{england2013statistical, england2015dissipative, perunov2016statistical}.  This organizing principle has been borne out in model systems \cite{kachman2017self,horowitz2017spontaneous}, yet has also been shown to have certain counter examples \cite{baiesi2018life}; it remains actively debated.  In the case of actomyosin systems, it has not yet been tested because of the difficulty in measuring dissipated free energy using most experimental approaches.  In this paper we take the first steps toward such a test, by developing a simulation methodology allowing the quantification of dissipated free energy during the self-organization of actomyosin networks.

In addition to being of interest in the field of active matter systems, dissipation in actomyosin networks has also been an important factor in recent experimental studies of cell mechanics.  Rheological properties of actomyosin networks largely determine rheological properties of the whole cell, and it has been discovered that the dissipative component of the cell's viscoelastic response to mechanical oscillations (called the loss modulus) is partly attributable to the ATPase activity of myosin motor filaments in actomyosin cytoskeleton \cite{balland2005dissipative, mitrossilis2009single, hoffman2006consensus}.  In the context of cell mechanics, myosin filaments have several roles: they produce mechanical stress by pulling on actin filaments, they dissipate mechanical stress by disassembling actin filaments and higher order stress-sustaining filament structures, and they consume chemical energy through the hydrolysis of ATP molecules.  It has been proposed that a lack of detailed understanding of the effects of these processes, and of dissipation in actomyosin systems more generally, underlies inconsistent, widely variable traction force microscopy measurements of cell migration \cite{kurzawa2017dissipation}.  Progress along this line is hindered by an absence of methods to study dissipation in actomyosin networks directly and at sufficiently high spatio-temporal resolution.  

To address these needs, we introduce a computational approach to measure dissipation during simulations of actomyosin network self-organization using the simulation platform MEDYAN (Mechanochemical Dynamics of Active Networks) \cite{popov2016medyan, hu2010mechano}.  MEDYAN simulations marry stochastic reaction-diffusion chemistry algorithms with detailed mechanical models of actin filaments, cross-linkers, myosin motor filaments, and other associated proteins, and it also accounts for mechanosensitive reaction rates.  This combination of simulation features makes this software uniquely capable of probing the complexity of actomyosin network dynamics.  For instance, past studies utilizing MEDYAN have investigated the dependence of network collapse on myosin filament and cross-linker concentrations, as well as the origin of local contractility in actomyosin networks \cite{popov2016medyan, komianos2018stochastic}.  We refer the reader to the paper describing MEDYAN for a detailed discussion of the various aspects of the simulation platform \cite{popov2016medyan}, while here we describe an extension of that platform that allows for calculation of the energetics of the chemical and mechanical events occurring during simulation.  We utilize these new capabilities to characterize the dissipation resulting from filament treadmilling, for which we further introduce a mean-field model, as well as from myosin filament walking.  We study both the time-dependence and the distributions of dissipation rates as concentrations of cross-linkers and myosin filaments are varied, observing that transduction of chemical energy to stored mechanical energy is more efficient at denser network organizations.  For these simulations, we first explore systems with ``plain" myosin filaments and cross-linkers that are not mechanosensitive, in order to simplify the overall dynamics.  We then introduce their mechanochemical coupling to understand its effect on the observed trends.  We end by discussing how this new methodology can provide a valuable technique to advance the studies of actomyosin networks mentioned above.

\section{Methods}
\subsection{Measuring Dissipation in MEDYAN}
We first give a brief overview of the MEDYAN simulation platform.  MEDYAN employs a stochastic chemical evolution algorithm in conjunction with mechanical representations of polymers and cross-linking proteins to simulate the dynamics of networks with active components, including but not limited to actomyosin networks.  The simulation space comprises a grid of reaction-diffusion compartments, inside which chemical species (e.g. unpolymerized subunits or cross-linking proteins) are assumed to be homogeneously distributed without specified locations, and which participate in reactions (e.g. (de)polymerization or (un)binding) according to mass-action kinetics; the species can additionally jump between compartments in diffusion events.  When an unpolymerized subunit polymerizes to or nucleates a filament, it becomes part of the mechanical subsystem, gaining location coordinates in the simulation volume and becoming subject to mechanical potentials depending on its interaction with other mechanical elements.  Through chemical reactions such as myosin filament binding and walking, the mechanical energy of the system changes, and the new net forces are then periodically relaxed in a mechanical equilibration phase, using conjugate gradient minimization.  We fill in salient details of the above overview as they become relevant below.  A user provides input data including system size and simulation length, mechanical parameters (e.g. stretching and bending constants and excluded volume cutoff distances), size of the polymer subunits, energy minimization algorithm parameters, chemical simulation algorithm parameters, choices for the modeling of force-sensitive reaction rates, initial conditions of the filaments (either specified or randomly generated), a list of reacting species and their associated parameters, a list of reactions involving those species and their associated parameters, and a list of desired output information.  The output of a simulation is a set of trajectory files containing information at each time point, which can include positions of the mechanical network elements, tensions on the elements, and copy numbers of the chemical species, among other things.  In Supplementary Information we discuss parameterization of the simulations analyzed in this paper.  MEDYAN is extensible in that it is possible to implement new types of outputs, depending on the experimental needs; in this paper we describe a novel output that reports the changes in the Gibbs free energy of the system.

As a MEDYAN simulation progresses, the Gibbs free energy of the system continually changes due to occurrences of chemical reactions and structural rearrangements of the polymer network.  These processes are driven by an out-of-equilibrium concentration of ATP which fuels filament treadmilling and myosin filament walking.  Dissipation measurement in MEDYAN works by calculating running totals of the chemical and mechanical energy changes. The running totals can then be converted into instantaneous rates by taking the numerical derivative at each time point using the forward difference quotient.  The algorithm for tracking these energy changes is compatible with the following sequence of consecutive procedures that make up one iterative cycle of a MEDYAN simulation \cite{popov2016medyan}:
\begin{enumerate}
	\setlength\itemsep{0em}
	\item Evolve system with stochastic chemical simulation for time $t_\text{min}$.  
	\item Calculate the changes in the mechanical energy resulting from the reactions in Step 1.
	\item Mechanically equilibrate the network based on the new stresses calculated in Step 2.  
	\item Update the reaction rates of force-sensitive reactions based on the new forces.
\end{enumerate}      

\begin{figure}[H]
	\centering
	\includegraphics[width= 15 cm]{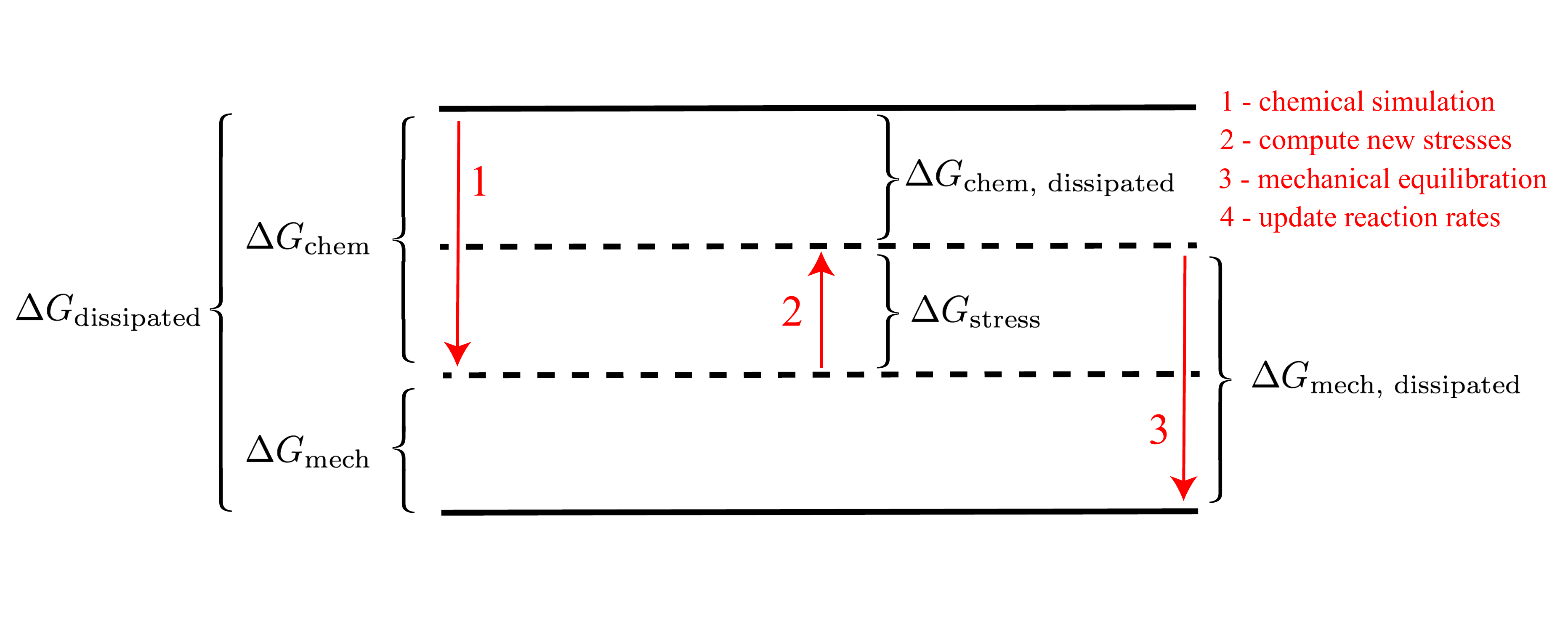}
	\caption{Energy level diagram indicating which changes in free energy are tabulated during the 4 procedures constituting one cycle in MEDYAN simulation.  Step 4, mechanochemical update of reaction rates, does not result in a free energy change.  Dotted lines represent energy levels that are intermediate during the iterative cycle, and solid lines represent the energies at the beginning and end of one cycle.}
	\label{enlevels}
\end{figure}

Dissipation tracking is done by calculating for each of these four steps a change in free energy, and then adding these free energy changes to determine the total change in free energy resulting from each cycle, $\Delta G_\text{dissipated}$.  Since, at least in this study, the actomyosin network is not mechanically coupled to any work reservoir external to the simulation volume (i.e. $W=0$), $\Delta G_\text{dissipated}$ is indeed dissipated energy \cite{ragazzon2018energy}.  This methodology could be straightforwardly extended to account for work exchanged with an external system in future studies, however.  Step 4 in the MEDYAN simulation cycle does not result in a change in free energy, as explained in Supplementary Information.  The notation chosen for these free energy changes, as well as the direction in which the energy is changing during each procedure, is illustrated in Figure \ref{enlevels}.

From this picture, we have the following relations:
\begin{align}
	\Delta G_\text{dissipated} &= \Delta G_\text{chem} + \Delta G_\text{mech}  \label{eq1} \\
	\Delta G_\text{dissipated} &= \Delta G_\text{chem, dissipated} + \Delta G_\text{mech, dissipated} \label{eq2} \\
	\Delta G_\text{chem, dissipated} &= \Delta G_\text{chem} + \Delta G_\text{stress} \label{eq3} \\
	\Delta G_\text{mech, dissipated} &= \Delta G_\text{mech} - \Delta G_\text{stress} \label{eq4}
\end{align}
For the depicted relative position of energy levels, the sign convention is such that all values of $\Delta G$ except for $\Delta G_\text{stress}$ will be negative (indicated by the arrow's direction), since $G$ refers to the free energy of the system, not of its environment, and will therefore be tend to be negative as the system moves down the free energy landscape.  The usual intuition that the total dissipation is positive can be stated 
\begin{equation}
\Delta G^+_\text{dissipated} = - \Delta G_\text{dissipated} >0 \label{eq5},
\end{equation}
where the superscript ``+" indicates the positive change in the total entropy.

Equation \ref{eq3} says that, given some change in the system's chemical potential energy, $\Delta G_\text{chem}$, resulting from reactions occurring during Step 1, a portion of that energy is used to deform the polymer network (e.g. via myosin filament pulling on actin filaments).  This increases the mechanical energy in the network by an amount $\Delta G_\text{stress}$.  Only the portion of $\Delta G_\text{chem}$ which has not gone into $\Delta G_\text{stress}$ has been dissipated as heat.  In Step 3 the network is mechanically equilibrated, resulting in relaxation of net forces (though not of all stresses) and updating of the network elements' positions.  We refer to the decrease in mechanical energy resulting from this relaxation as $\Delta G_\text{mech, dissipated}$.  

The calculation of $\Delta G_\text{stress}$ and $\Delta G_\text{mech, dissipated}$ is based on a set of mechanical potentials describing interactions between elements of the actomyosin network.  Polymers are modeled as a sequence of thin, unbendable, yet extensible cylinders that are joined at their ends by beads whose positions define the polymer's configuration.  The structural resolution of MEDYAN is at the level of the cylinders, which in this study are 27 $nm$ long and have effective diameters of approximately 5 $nm$, however the diameter is not a parameter of the simulation, being instead effectively determined by the strength of the excluded volume interaction between cylinders.  Cross-linking proteins (e.g. $\alpha$-actinin and myosin filaments) are modeled as Hookean springs connecting these cylinders by attaching to discrete binding sites.  Included among the mechanical potentials are various modes of filament deformation, excluded volume interactions, and stretching of cross-linkers and myosin filaments.  Mechanical equilibration is accomplished by constrained minimization of the mechanical energy with respect to the positions of the network elements.   A full description of the mechanical potentials and equilibration protocols is given in \cite{popov2016medyan}.  Determining $\Delta G_\text{stress}$ and $\Delta G_\text{mech, dissipated}$ requires evaluating the instantaneous total mechanical energy of the system at certain points during the iterative simulation cycle and taking the difference of those values.  

The calculation of $\Delta G_\text{chem}$ and $\Delta G_\text{chem, dissipated}$ is accomplished by incrementing a running total of the chemical energy $G_\text{chem}$ whenever a reaction stochastically occurs during Step 1, and finding the accumulated change at the end of the protocol.  Chemical stochastic simulation in MEDYAN uses a Gillespie-like reaction-diffusion algorithm over a grid of compartments which constitutes the simulation volume.  Diffusing species are assumed to be homogeneous (i.e. obey mass-action kinetics) inside the compartments, and can jump between the compartments leading to concentration gradients at the scale of the compartment length (taken to be roughly the Kuramoto length of diffusing G-actin, following \cite{hu2010mechano}).  The evolving polymer network is overlaid on this compartment grid, with each piece of a polymer reacting with diffusing species according to the concentrations in its local compartment.  Again, we refer the reader to \cite{popov2016medyan} for a more detailed description of the chemical dynamics.  For the present purpose of measuring dissipation, we introduced into this simulation protocol a precise formula for the change in Gibbs free energy corresponding to the occurrence of various reactions as a function of the instantaneous compartment concentrations.  In the Supplementary Information we establish this formula, while we present its derivation in an accompanying paper \cite{2019arXiv190110520F}.  The set of chemical reactions used to describe actomyosin networks in this study is based on a previous model of actin polymerization dynamics that explicitly treats hydrolysis states of the nucleotide bound to each actin subunit \cite{brooks2009nonequilibrium, floyd2017low}.  This level of detail allows to quantify the dissipation resulting from ATP hydrolysis during filament treadmilling.  To increase computational efficiency, we neglect the dynamics of nucleotide hydrolysis states of the tips of the filaments.  This has been shown in previous work to be a valid approximation to the full dynamics which includes the states of the filament tips \cite{floyd2017low}.  We refer to the resulting set of reactions describing actin polymerization dynamics as the Constant Tip (CT) model.  However unlike in the original CT model, here we explicitly include G-actin bound to ADP-Pi as a reacting species, for completeness and since the extra computational strain of doing so is small.  The actin subunit species tracked in this model are distinguished by their polymerization state and by the hydrolysis states of the nucleotide to which they are bound.  We notate a species as $G$ or $F$, to represent globular (un-polymerized) or filamentous (polymerized) actin respectively, superscripted by $T$, $Pi$, or $D$ to represent that it is bound to ATP, ADP-Pi, or ADP, respectively; thus for instance filamentous actin bound to ADP-Pi is notated $F^{Pi}$. We also include reactions describing cross-linker (un)binding and myosin filament (un)binding and walking.  We exclude filament nucleation, severing, destruction, and annealing reactions, thus the number of filaments is constant throughout the simulation trajectories.  In the Supplementary Information we describe how we compute the change in Gibbs free energy for each reaction in this set, as well as how we parameterize the simulations.

Lastly, we developed a mean-field model to describe just the dissipation resulting from reactions in the CT model, i.e. excluding cross-linkers and myosin filaments.  We describe the model and present its results in Supplementary Information.  We find, among other things, that the steady-state dissipation rate from filament treadmilling counter-intuitively does not depend on the total amount of actin, but only on the number of filaments.

\section{Results}

\subsection{Total Dissipation Rates of Disordered Networks Do Not Increase}

\begin{center}
	\begin{figure}[H]
		\centering
		\includegraphics[width= 17 cm]{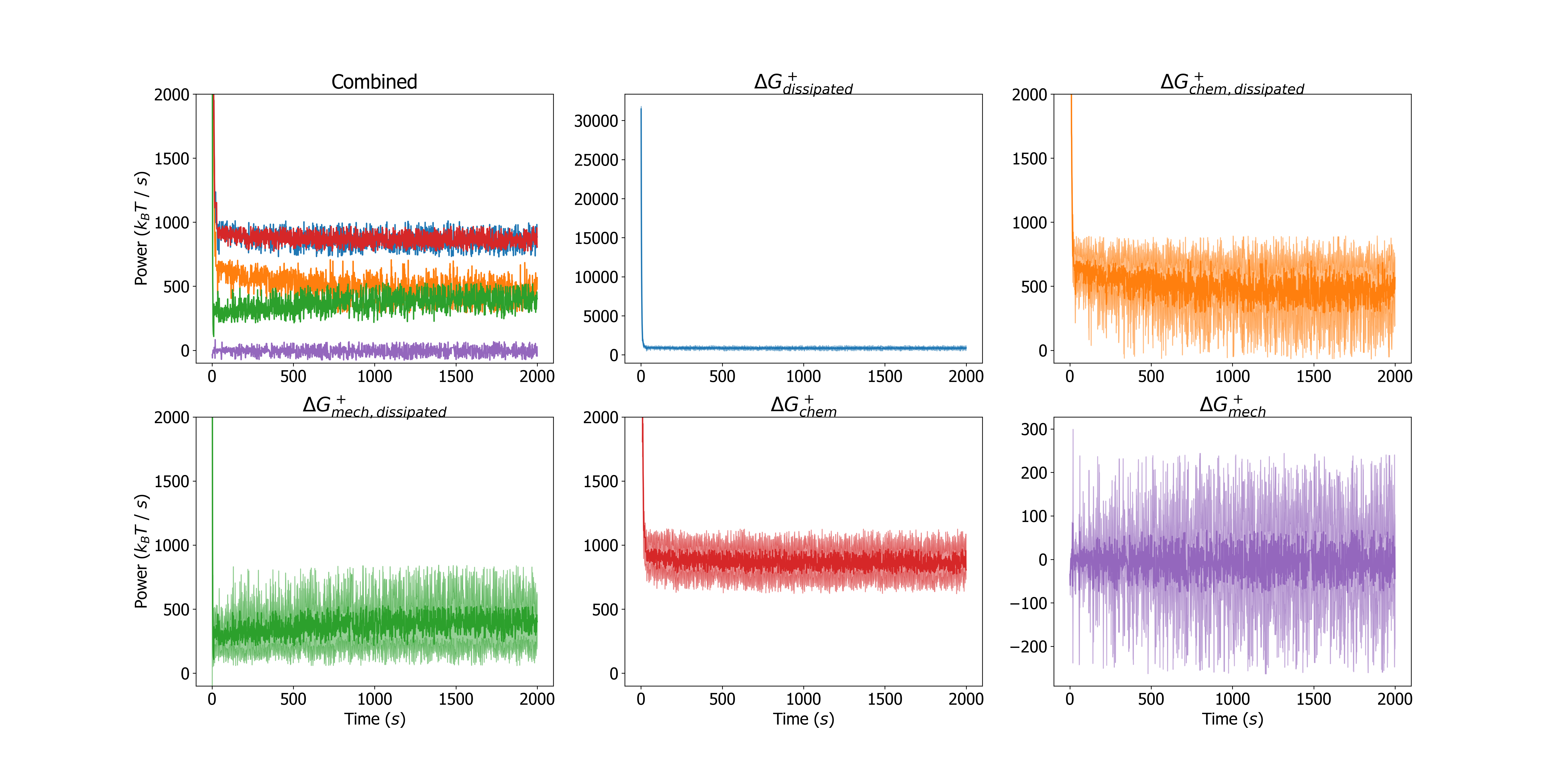}
		\caption{\textit{Top Left}: Combined trajectories of 5 quantities tracked during a MEDYAN simulation, averaged over 10 separate runs.  The color coding is indicated by the remaining panels.  Note the close overlap between $\Delta G^+_\text{dissipated}$ and $\Delta G^+_\text{chem}$.  In the remaining panels, the individual trajectories are visualized with their standard deviations at each time point over the 10 runs visualized as lighter curves above and below the main curve.  In the plots for $\Delta G^+_\text{dissipated}$ and $\Delta G^+_\text{mech}$, the full range is visualized, however this range is cropped in the other plots to aid visibility.}
		\label{trajcomp}
	\end{figure}
\end{center}

We studied dissipation rates accompanying the process of myosin-driven network self-organization.  We first excluded in these simulations the force-sensitivity of the reaction rates describing cross-linker unbinding and myosin filament unbinding and walking.  This allowed us to understand a simplified version of the dynamics (we later discuss the the effect of including mechanochemical feedback).  We analyzed the trajectories of the quantities $\Delta G_\text{dissipated}$, $\Delta G_\text{chem, dissipated}$, $\Delta G_\text{mech, dissipated}$, $\Delta G_\text{chem}$, and $\Delta G_\text{mech}$ over a set of simulations with identical initial concentrations but different random filament distributions.  In Figure \ref{trajcomp}, we display these trajectories averaged over 10 runs.  We used a simulation volume of 1 $\mu m^3$, divided into 8 compartments, and initial conditions of equal amounts (10 $\mu M$ each) of $G^T$ and $G^D$ actin in a 0.08 $\mu M$ pool of seed filaments containing $F^T$,  as well as 0.2 $\mu M$ myosin and 1.0 $\mu M$ cross-linkers.  Similar trajectories for other conditions are shown in SI Figures \footnote{\label{dns}Due to file size constraints on preprint upload, this supplementary figure has been excluded.  It will be available in the print version coming soon in \textit{Interface Focus}.}.  Points lying outside 3 median absolute deviations (MAD) have been excluded for this visualization \cite{leys2013detecting} \footnote{For certain statistical aggregations in this paper we use the MAD since it is the estimate of scale most robust against outliers, having a breakdown point of 50\%.  However, we use the unscaled version of the MAD and, as a result, do not claim that this statistic is a consistent estimator of the standard deviation \cite{huber2011robust}.  The distributions of most of the quantities of interest are too pathological to allow for a straightforward choice of scaling factor, so we simply us the unscaled version, defined as $\text{MAD} = \text{median}_i  |x_i - \text{median}_jx_j | $, where $x_i$ are elements in the data set.}.

\begin{center}
	\begin{figure}[H]
		\centering
		\includegraphics[width = 13 cm]{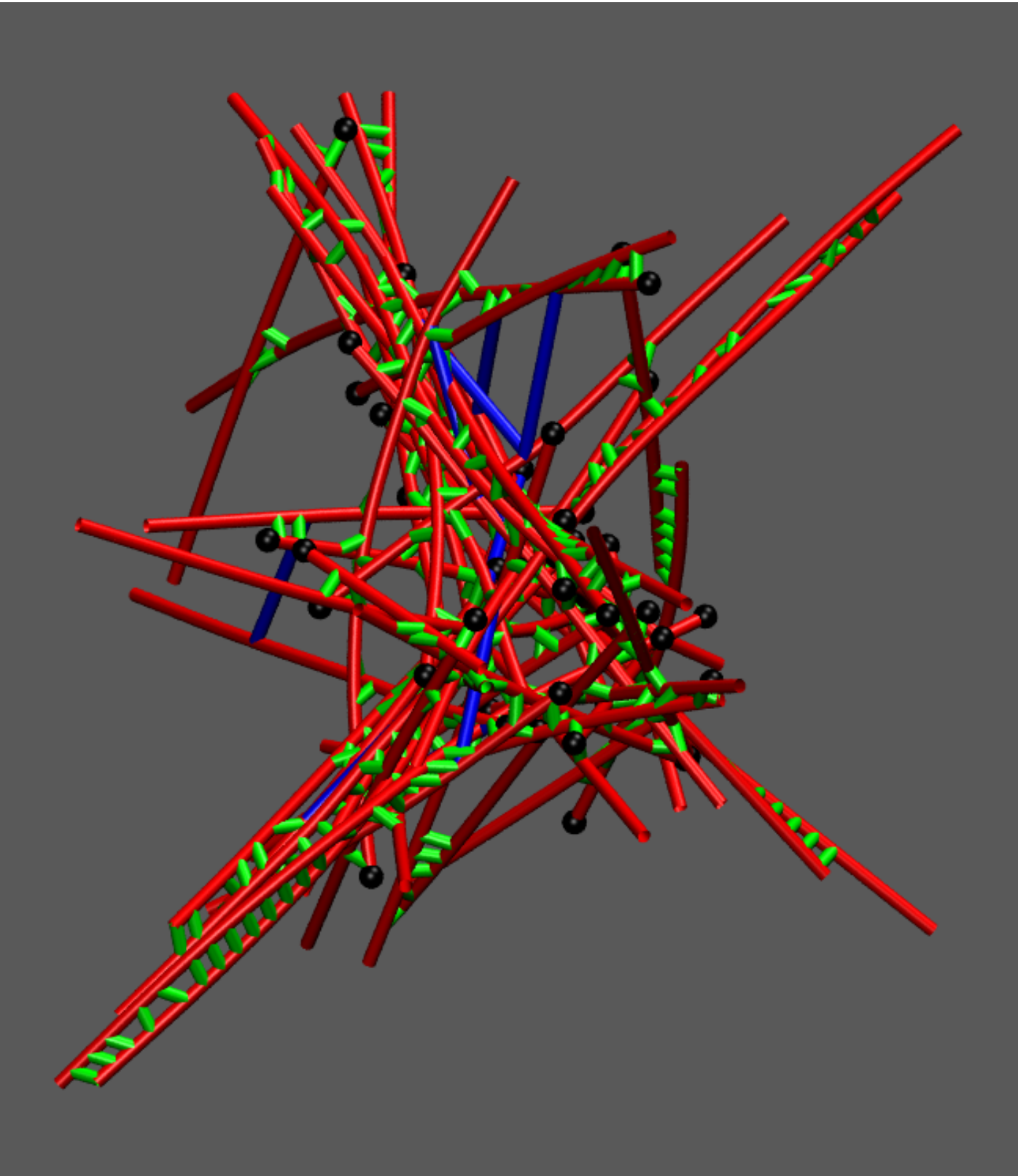}
		\caption{Snapshot of a percolated actomyosin network in MEDYAN under conditions $C_{CL}$ = 5.0 $\mu M$ and $C_M$ = 0.4 $\mu M$.  Actin filaments are drawn in red, cross-linkers are drawn in green, myosin filaments are drawn in blue, and the plus ends of filaments are drawn as black spheres.  }  
		\label{hhsnap}
	\end{figure}
\end{center}

\begin{center}
	\begin{figure}[H]
		\centering
		\includegraphics[width = 9 cm]{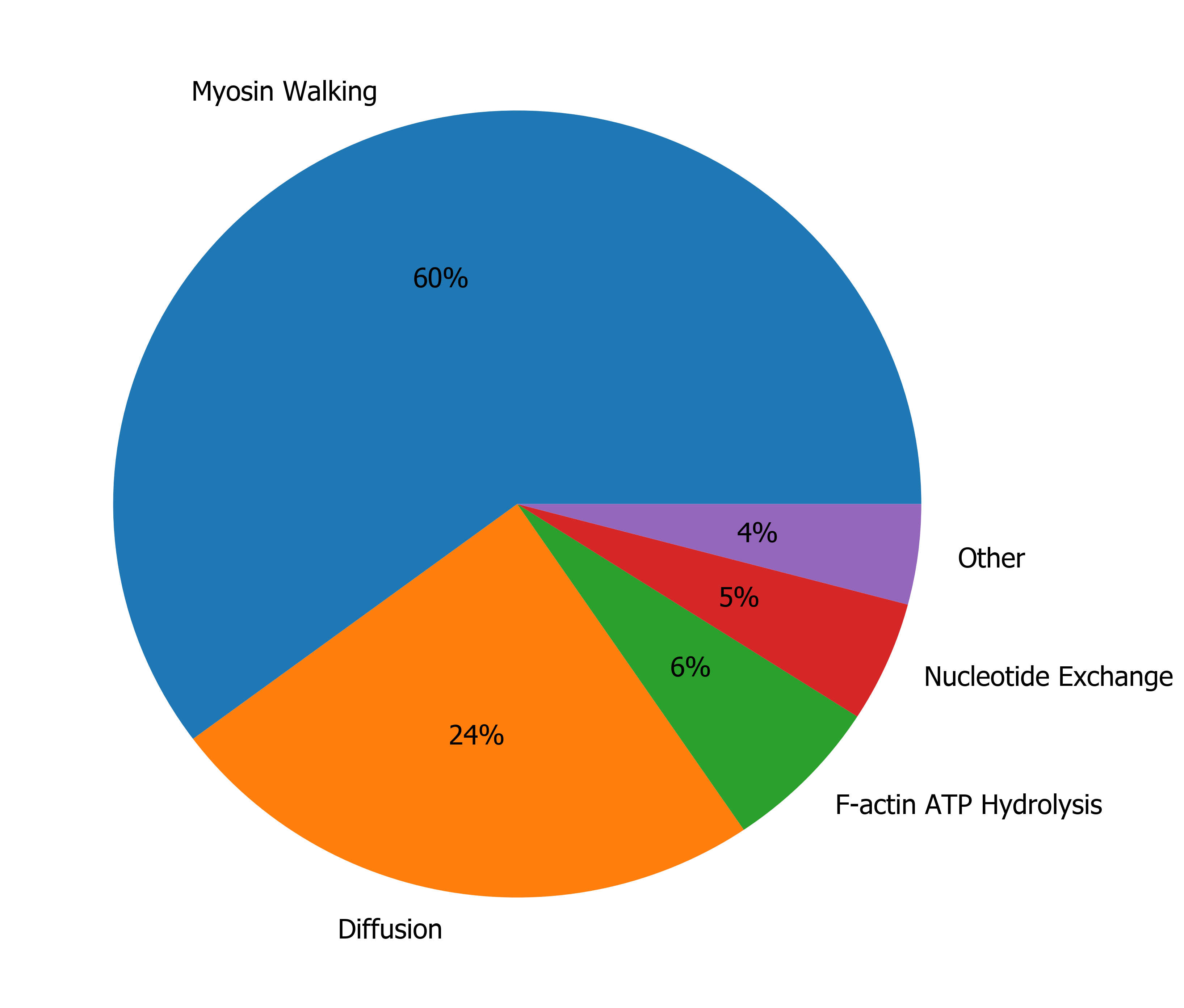}
		\caption{Integrated contributions of each reaction in MEDYAN to the total $\Delta G_\text{chem}$ along a simulation trajectory with $C_{CL}$ = 1.0 $\mu M$ and $C_M$ = 0.2 $\mu M$.}
		\label{piemm}
	\end{figure}
\end{center}

For each tracked quantity, there is an initial transient phase (lasting a few tens of seconds) followed by fluctuations around a roughly steady value; we do not observe in this set of simulations any slow approach to a significantly different total dissipation rate, which might correspond to large reorganizations of the network.  However for high concentrations of cross-linkers ($C_{CL} = 5.0$ $\mu M$) and myosin ($C_{M} = 0.4$ $\mu M$), we observe that the contribution of $\Delta G^+_\text{mech, dissipated}$ to $\Delta G^+_\text{dissipated}$ tends to increase relative to that of $\Delta G^+_\text{chem, dissipated}$ (SI Figure \footref{dns}).  It is known that network percolation and collapse occurs only under certain conditions of myosin and cross-linker concentrations \cite{popov2016medyan, alvarado2013molecular}, and it is under these conditions which result in collapse that we observe an increase in $\Delta G^+_\text{mech, dissipated}$ relative to $\Delta G^+_\text{chem, dissipated}$ (SI Figures \footref{dns}).  This indicates that more mechanical stress is being created by myosin filament walking as the network collapses and becomes more densely cross-linked. The transient phase corresponds to the initial polymerization of the seed filaments followed by the initial mechanical coupling of filaments by cross-linkers and myosin filaments.  Following the transient phase, the networks in this study are generally disordered (Figure \ref{hhsnap}).  The dissipation rate corresponding to the initial polymerization of the seed filaments is much larger than the chemical dissipation resulting from myosin filament activity.  However, following the transient phase, after which the treadmilling dissipation has reached its steady-state (SI Figure \ref{combined}), the contribution from myosin filament activity outweighs the dissipation resulting from filament treadmilling.  Tracking the instantaneous rate of change in $\Delta G^+_\text{chem}$ resulting from each reaction separately, we observe myosin filament walking to contribute the majority to $\Delta G^+_\text{chem}$ after the transient phase, however the amount depends on $C_{CL}$ and $C_M$.  The integrated contributions of each reaction to $\Delta G^+_\text{chem}$ for different conditions are shown in Figure \ref{piemm} and SI Figures \footref{dns}.  Interestingly, diffusion contributes an appreciable fraction to $\Delta G^+_\text{chem}$; plus and minus ends of actin filaments tend to localize together (Figure \ref{hhsnap}) and through treadmilling deplete the concentrations of $G^T$ and $G^D$ in certain reaction compartments relative to others, leading to significant diffusion gradients on the scale of the compartment length.  This suggests that the establishment of concentration gradients is an important driving force in actomyosin self-organization, as has been noted in other work \cite{erban2014multiscale, hu2013molecular}.  In the initial transient phase, the mechanical energy changes appreciably as the filaments grow and are initially coupled to each other.  Following this, however, the rate of $\Delta G^+_\text{mech}$ is on average near zero.  This indicates that, despite the process of mechanical stress generation through myosin filament activity and treadmilling, the resulting stress is dissipated through fast relaxation such that, on a slower timescale, the mechanical energy of the system does not change in a significant, persistent way.     

\begin{center}
	\begin{figure}[H]
		\centering
		\includegraphics[width= 17 cm]{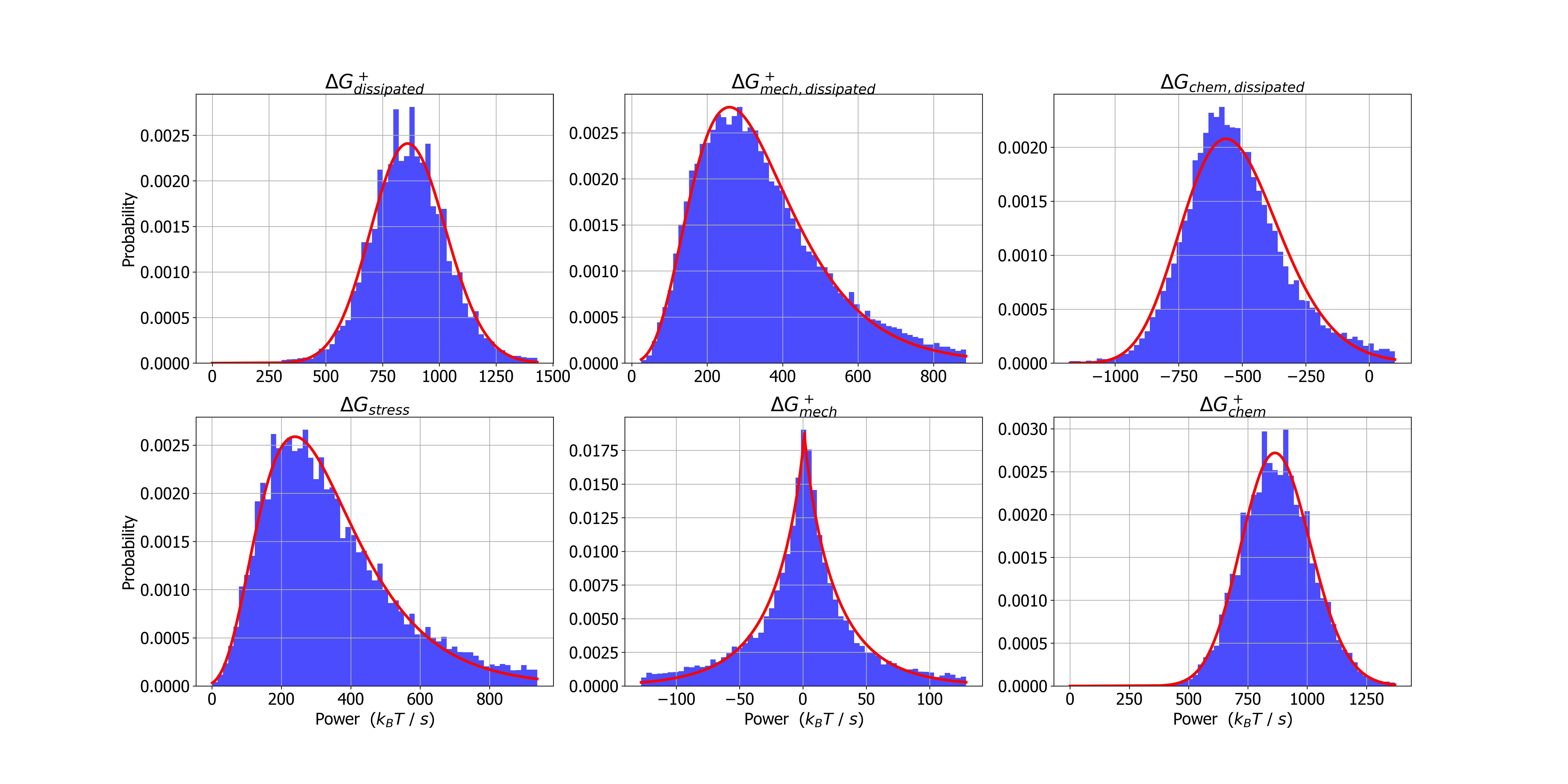}
		\caption{Histograms and fitted probability distribution functions for 6 tracked quantities.  For each histogram, the full trajectory for each of 10 runs is combined into a single data set.  A log-normal distribution was used to fit the histograms of $\Delta G^+_\text{dissipated}$ and $\Delta G^+_\text{chem}$, a generalized normal distribution was used to fit $\Delta G^+_\text{mech}$, and the rest were fit with gamma distributions.  All distributions are fit using the SciPy package to determine shape, scale, and location parameters \cite{jones2014scipy}.  Quantities were made positive or negative in order to produce the best fits.}
		\label{histcomp}
	\end{figure}
\end{center}

Aggregating each of the 2000 $s$ of the 10 trajectories into a collective data set for each condition of $C_M$ and $C_{CL}$, we next analyzed the distribution of the instantaneous rates of $\Delta G^+_\text{dissipated}$, $\Delta G^+_\text{mech, dissipated}$, $\Delta G^+_\text{chem, dissipated}$, $\Delta G^+_\text{stress}$, $\Delta G^+_\text{mech}$, and $\Delta G^+_\text{chem}$.  In Figure \ref{histcomp} we plot histograms of these 6 quantities for the conditions $C_{CL} = 1.0$ $\mu M$, $C_{CL} = 0.2$ $\mu M$.  In SI Figures \footref{dns} we show the same plots for other conditions.   

Each distribution contained heavy tails which, for the purpose of fitting, we suppressed by excluding data lying outside 5 MAD's of the median.  No distributions were sufficient to cleanly fit the full set of data, so we focus here on the center of the distribution and include a qualitative discussion of the heavy tails below. With the exception of $\Delta G^+_\text{mech}$ which was fit with a generalized normal distribution, each distribution exhibited significant skew and could be fit reasonably well with a log-normal or a gamma distribution.  At higher concentrations of myosin and cross-linkers, the histograms were less cleanly fit by any standard distributions (SI Figure \footref{dns}).

Log-normal distributions are fairly ubiquitous across different fields and systems \cite{limpert2001log}.  For instance, it has been shown that the distributions of concentrations of species in a chemical reaction network are log-normal  \cite{furusawa2005ubiquity}.  Gamma distributions are similarly common and often difficult to discriminate from log-normal distributions \cite{kundu2005discriminating}.  A speculative explanation for the gamma distribution of $\Delta G_\text{stress}$ is as follows: $\Delta G_\text{stress}$ can be viewed as resulting from a number of myosin filament steps that, according to the central limit theorem, is approximately normally distributed given a sufficiently long time between simulation snapshots $t_\text{snap}$ (these stepping events are not truly i.i.d., but to a first approximation we may assume they are).  The main effect of each of these steps is to increase the harmonic stretching potential on the myosin filaments as well as on the actin filament cylinders by a roughly fixed stepping distance.  Thus the increase in mechanical energy, $\Delta G_\text{stress}$, is approximately a quadratic function of the normally distributed number of myosin filament steps per $t_\text{snap}$.  As shown in SI Figure \footref{dns}, the resulting distribution of $\Delta G_\text{stress}$ is well-fit by a gamma distribution and bears qualitative similarity to the histogram of $\Delta G_\text{stress}$ in Figure \ref{histcomp}.  In Figure \ref{histcomp}, $\Delta G_\text{stress}$ and $\Delta G^+_\text{mech, dissipated}$ have similar distributions, indicating that, for each MEDYAN cycle, almost all the stress accumulated following Step 1 is then immediately relaxed.  The remainder goes into $\Delta G^+_\text{mech}$, whose distribution is centered on zero with little skew.  Furthermore,  the distribution of $\Delta G^+_\text{mech, dissipated}$ has particularly heavy tails, indicating infrequent yet large relaxation events.  This tendency has some precedent in avalanche-prone systems, whose hallmarks are self-organized criticality, intermittency, and scale-invariance in their distribution of avalanche event sizes \cite{bak1988self, bak1989earthquakes, sornette1989self, bottiglieri2007off}.  While we do not observe true scale-invariance in this set of simulations (i.e. power-laws cannot fit these distributions cleanly), it will be interesting to continue to explore actomyosin networks dynamics in the framework of self-organized criticality \cite{cardamone2011cytoskeletal}.

\subsection{More Compact Networks are More Efficient}
\begin{center}
	\begin{figure}[H]
		\centering
		\includegraphics[width= 15 cm]{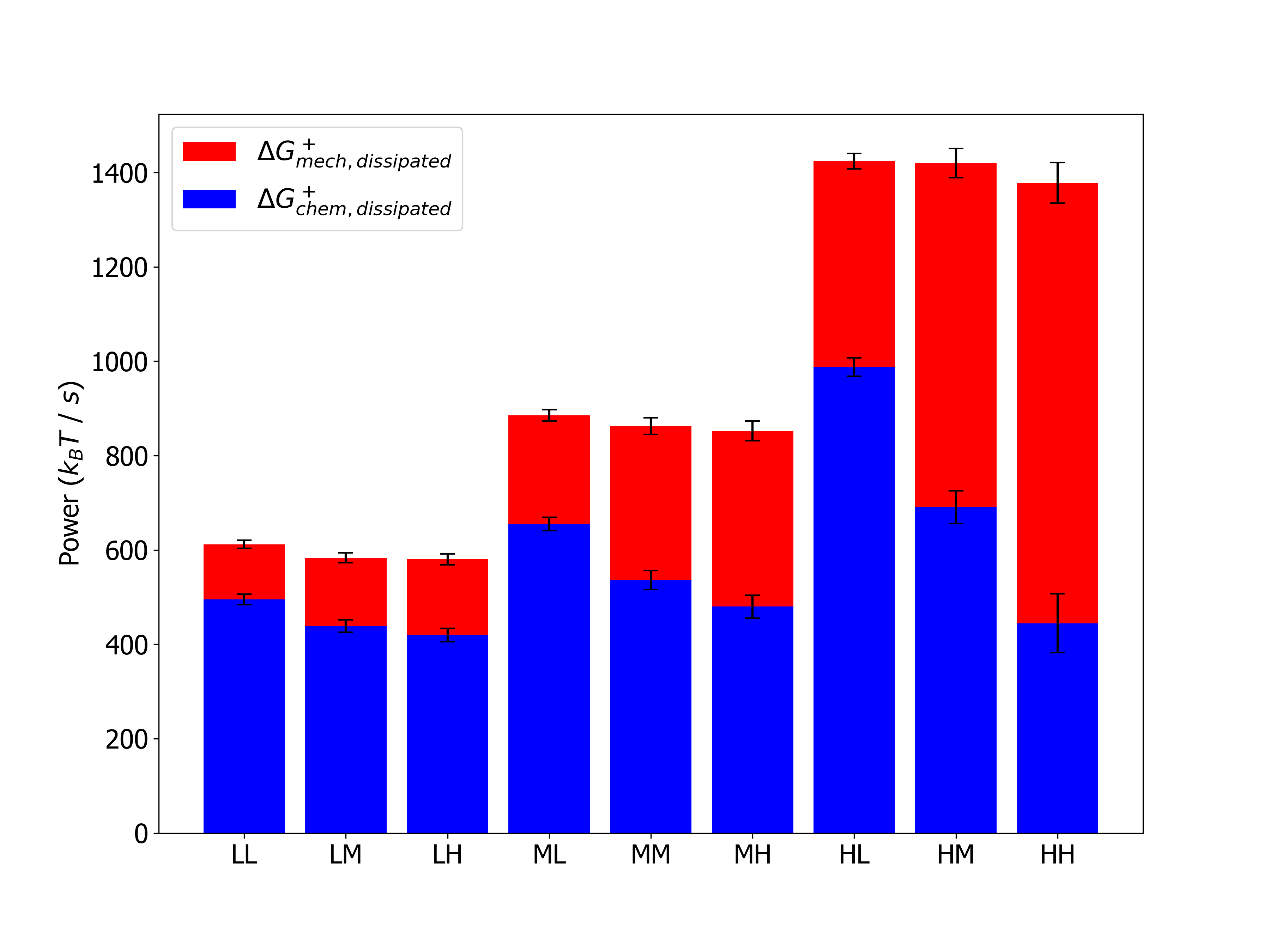}
		\caption{Bar plot representing the contributions of $\Delta G^+_\text{chem, dissipated}$ and $\Delta G^+_\text{mech, dissipated}$ to the total, $\Delta G^+_\text{dissipated}$.  The letters in the abscissa labels designate ``low," ``medium," and ``high."  The first letter represents the concentration of myosin, $C_{M}$: ``L" = $0.1 \ \mu M$, ``M" = $0.2 \ \mu M$, ``H" = $0.4 \ \mu M$, and the second letter represents the concentration of cross-linkers, $C_{CL}$: ``L" = $0.1 \ \mu M$, ``M" = $1.0 \ \mu M$, ``H" = $5.0 \ \mu M$.   The median of each quantity is taken over 10 runs of 2000 $s$, and error bars represent 1 MAD. }
		\label{barplot}
	\end{figure}
\end{center}

We simultaneously varied concentrations of myosin filaments $C_M$, and cross-linkers, $C_{CL}$, which is known to produce a range of network architectures \cite{popov2016medyan, freedman2018nonequilibrium}.  Using cross-linker concentrations of 0.1, 1.0, and 5.0 $\mu M$, and myosin concentrations of 0.1, 0.2, and 0.4 $\mu M$, we studied the effects on $\Delta G^+_\text{dissipated}$, $\Delta G^+_\text{chem, dissipated}$, and $\Delta G^+_\text{mech, dissipated}$.  The concentrations of actin subunits and filaments are the same as described above.  For each of the 9 conditions, we ran 10 simulations of 2000 $s$.  In Figure \ref{barplot}, we display the median values of $\Delta G^+ _\text{chem, dissipated}$ and $\Delta G^+ _\text{mech, dissipated}$ along each trajectory and over each repeated trajectory for these conditions.  The median is used here because of its insensitivity to outliers such as are found in the heavy-tailed distributions of these quantities \cite{huber2011robust}.  
  
We find that as $C_{M}$ is increased, the median rates of $\Delta G^+_\text{dissipated}$, $\Delta G^+_\text{chem, dissipated}$, and $\Delta G^+_\text{mech, dissipated}$ all tend to increase.  Further, the value of of $\Delta G^+_\text{mech, dissipated}$ relative to $\Delta G^+_\text{chem, dissipated}$ increases.  As $C_{CL}$ is increased with $C_{M}$ fixed, then $\Delta G^+_\text{chem, dissipated}$ tends to decrease.  Increased concentrations of myosin filaments obviously have a strong effect on the dissipation simply because they are the chief active agents in the system once treadmilling has reached its steady-state.  We can define a measure of efficiency for the present purpose as:

\begin{equation}
\eta = \frac{\Delta G_\text{stress}}{\Delta G^+_\text{chem}} = \frac{\Delta G^+_\text{mech, dissipated} - \Delta G^+_\text{mech}}{\Delta G^+_\text{dissipated} - \Delta G^+_\text{mech}} \approx \frac{\Delta G^+_\text{mech, dissipated}}{\Delta G^+_\text{dissipated}}
\label{eq24}
\end{equation}
where the approximation follows since, as illustrated in Figure \ref{trajcomp}, the average of $\Delta G^+_\text{mech}$ is close to zero.  Thus it is evident that, perhaps surprisingly, as $C_{M}$ increases, $\eta$ increases: the more motors are in our system, the more efficiently can chemical energy be converted into mechanical stresses.  Further, as $C_{CL}$ increases, $\eta$ tends to increase because $\Delta G_\text{chem, dissipated}$ decreases relative to $\Delta G_\text{mech, dissipated}$.  At higher levels of cross-linking, which introduce mechanical constraints on the filaments, the walking of myosin motor filaments will produce more stress compared to at low levels of cross-linking, when filament sliding can result from myosin filament walking, producing less stress.  

We repeated the above experiments with the inclusion of mechanochemical feedback on the reaction rates controlling cross-linker unbinding and myosin filament unbinding and walking  (SI Figures \footref{dns}).  We observed, somewhat surprisingly, no qualitative differences compared to the results described for the case of no feedback, however there were quantitative differences in the stress production and radius of gyration for certain conditions.  These quantitative differences result from the fact that, for the concentrations $C_M$ and $C_{CL}$ used above, we did not observe significant collapse of the actomyosin network when mechanochemical feedback was included due to the stalling and catching of the myosin filaments.  As a result the networks were less densely cross-linked, and the efficiency was lower.  The total dissipation rates are largely unaffected by the inclusion of mechanochemical feedback; it was primarily the degree to which chemical energy had been converted to stress that was different for certain conditions.

It is worthwhile to mention how these results compare with \textit{in vitro} studies of dissipation in actomyosin systems.  While the computational approach described in this paper allows uniquely highly-resolved and direct measurement of free energy changes, other experimental methods have produced qualitatively similar results to those obtained here.  Rheological experiments have determined that a single cell's response to compression is similar in nature to a muscle's response to increasing load, suggesting that the actomyosin network underlies the cell's mechanical responsiveness \cite{mitrossilis2009single}.  Further, this responsiveness is modulated by blebbistatin, a myosin ATPase inhibitor, highlighting myosin's role in negotiating how the network rearranges in response to the sustained stress.  Using the metric of mechanical dissipation to measure the degree of structural rearrangements and release of stress, we confirm that these processes indeed sensitively depend on myosin activity, which we control here through its concentration.  Additional rheological studies have probed the mechanical dissipation of actin cortices more directly, using the loss modulus as a readout.  These have also indicated that inhibiting myosin activity reduces the mechanical dissipation of the system, causing it to be behave more elastically \cite{balland2005dissipative}.  Lastly, we mention a recent study that quantified mechanical dissipation of actin filaments using a novel experimental method \cite{seara2018entropy}.  By measuring the flow through a low-dimensional phase space defined by the amplitudes of the filament bending modes \cite{battle2016broken}, they determine the entropy production of fluctuating actin filaments in different phases of contractility.  They relate the entropy production to the degree of transverse bending of the filaments, as opposed to sarcomeric filament sliding, caused by myosin filament walking.  Similarly, we here relate the hindrance of filament sliding due to cross-linker density to increased mechanical dissipation rates.  Quantitative comparisons of these two approaches to mechanical dissipation measurement will be an interesting future direction.  We note finally that a unique capability of quantifying dissipation using MEDYAN is the ability to simultaneously measure the energetics of chemical reactions in addition to the changes in the mechanical energy, which is not currently available using \textit{in vitro} methods.

\section{Discussion}

We have introduced a methodology for tracking the energetics of  chemical and mechanical events during a MEDYAN simulation, allowing us to probe the properties of actomyosin networks as dissipative active matter systems.  The distinction between dissipation's mechanical and chemical origins is natural in the context of MEDYAN's iterative simulation procedure which carries out chemical stochastic simulation, mechanical deformation, and mechanical relaxation at separate times.  As explained in \cite{popov2016medyan}, this procedure exploits a separation of timescales between the characteristic mechanical relaxation times of actomyosin networks \cite{falzone2015entangled} compared to typical waiting time between reactions that introduce mechanical stresses, such as myosin filament walking \cite{kovacs2003functional} or filament growth \cite{fujiwara2007polymerization}.  Ultimately the source of all dissipation is the chemical potential of ATP molecules driving treadmilling and myosin filament activity.  This is reflected by the near equality of $\Delta G^+_\text{chem}$ and $\Delta G^+_\text{dissipated}$ in Figure \ref{trajcomp} and SI Figures \footref{dns}.  On a fast timescale (that of the characteristic mechanical relaxation time), however, the free energy of chemical reactions cause small force deformations of the network which are then quickly and almost fully relaxed.  Thus, for a myosin filament stepping event, only a portion of the chemical free energy $\Delta G^+_\text{chem, dissipated}$ is immediately dissipated as heat, with the rest going into temporarily increased mechanical energy of the actomyosin network, $\Delta G_\text{stress}$.  The fast relaxation of this new mechanical stress constitutes what we refer to as mechanical dissipation, $\Delta G^+_\text{mech, dissipated}$, and the small residual stress after this relaxation has balanced all net forces acting on the system results in a change of the mechanical energy of the system on a slow timescale, $\Delta G^+_\text{mech}$.

\begin{center}
	\begin{figure}[H]
		\centering
		\includegraphics[width= 12 cm]{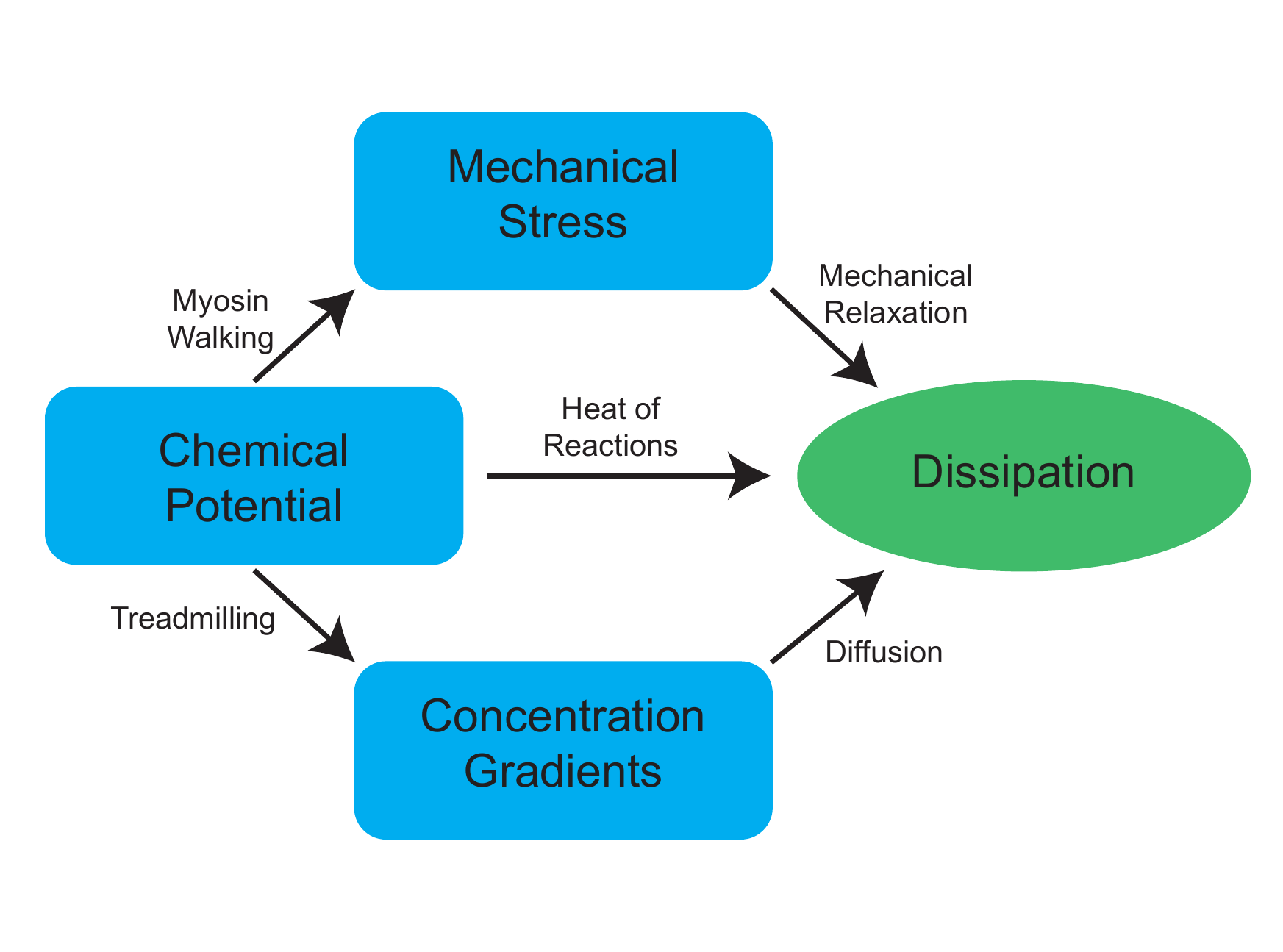}
		\caption{A simple schematic illustrating the flow of free energy in actomyosin network systems.  Blue compartments represent forms of free energy storage, arrows represent the transfer of free energy from one form to another, and the green compartment indicates  dissipation as ultimate destination of all free energy flows in this nonequilibrium system.  Arrows are labeled with the mechanisms by which these free energy transduction from one form to another are achieved.   In this depiction, the sizes of compartments and the widths of the arrows indicating the magnitudes of the represented quantities are not to scale. }
		\label{eflows}
	\end{figure}
\end{center}

One interpretive framework which is useful to understand the flow of free energy in actomyosin networks is illustrated in Figure \ref{eflows}.  We can think of the different forms of free energy storage, including chemical potentials, mechanical stress, concentration gradients (which could also be considered as arising from chemical potential differences across compartments), and dissipated energy, as nodes on a directed graph, where edges represent transduction of energy from one form of storage to another.  The weights of these edges represent the amount of free energy flowing through them.  In this picture, we can describe the process of network percolation, which occurs at increasing concentrations of cross-linkers and myosin filaments, as widening the edge flowing from chemical potential into mechanical stress, while thinning the edge from chemical potential directly to dissipation.  The edge weights corresponding to the establishment of concentration gradients and the resulting diffusive dissipation will not be affected dramatically by the onset of percolation except to the extent that percolated networks might lead to the formation of more bundles, and therefore more significant concentration gradients.  At steady state, we have a stationary current on the graph, fueled by the chemical potential of the assumed limitless supply of ATP.  Of course, at this level of description, we have coarse-grained away the details of the specific chemical reaction networks and mechanical potentials which constitute the system, however by doing so we gain a clearer understanding of how percolation of the network alters the flows of free energy.   

The new capabilities of MEDYAN allowing detailed energetics computations should provide a way to address outstanding issues in the the fields of active matter systems and of cell mechanics.  In this study we observe dissipation rates to stay at a low fluctuating steady-state following a high initial transient phase, apparently in contrast with the predictions of the dissipation-driven adaptation hypothesis.  However we also do not observe slow reorganization of the actomyosin network into different higher order structures such as bundles (for the concentrations of components used here we observe disordered networks only), and as a result we can not rule out that such reorganizations correspond to marked changes in the dissipation rates.  A dedicated test of the hypothesis of dissipative-driven adaptation should be straightforward with this methodology.  For instance, a simulated gliding assay, which has been shown \textit{in vitro} to lead to diverse dynamical patterns \cite{schaller2010polar}, may indicate that these emergent patterns correspond to an optimization of the chemical dissipation from myosin filament walking, as argued in \cite{england2015dissipative}.  In the context of cell migration, studies investigating the mechanically dissipative activity of myosin filaments are also feasible if one incorporates in simulation an external substrate against which the actomyosin network pulls.  In this type of study, it should be straightforward to determine how the amount of stress which is sustained against the substrate is altered by myosin activity given the energetics calculations in the methodology described in this paper.  It is also feasible to do simulated measurements of the dynamic shear modulus by compressing the actomyosin network at different frequencies.  It should then be possible to directly observe the degree to which the dissipation of elastically stored energy is attributable to myosin walking.  In fact, this last research question has already been investigated to some extent using an alternate computational model to the one presented here \cite{mcfadden2017filament}.  We hope some of these potential future experiments will shed light on outstanding questions in the studies of actomyosin networks.  

\begin{center}
\section{Supplementary Information for ``Quantifying Dissipation in Actomyosin Networks" }
\end{center}

	\subsection{$\Delta G$ of Chemical Reactions}

	For a reaction of the general form 
	\begin{equation}
		\label{eq6}
		\nu_1 X_1 + \nu_2 X_2 + \ldots \rightleftharpoons \upsilon_1 Y_1 + \upsilon_2 Y_2 + \ldots 
	\end{equation}
	we define $X_i$ as the reactant species, $Y_i$ as the product species, and $\nu_i$ and $\upsilon_j$ as their stoichiometric coefficients.  The ``stoichiometric difference" is defined as
	\begin{equation}
		\label{eq7}
		\sigma = \sum_{j \in P} \upsilon_j - \sum_{i \in R} \nu_i 
	\end{equation} 
	where $P$ is the set of products and $R$ is the set of reactants.
	We further define the conversion factor between the copy number of species $i$, $N_i$, and its concentration $C_i$,
	\begin{equation}
		\label{eq8}
		\Theta = N_\text{Av} V
	\end{equation}
	where $N_\text{Av}$ is Avogadro's number, $V$ is the volume of the compartment where the reaction occurs, and we have $N_i = C_i \Theta$.  Next we define the following quantity, reminiscent of the reaction quotient:
	\begin{equation}
		\label{eq9}
		\widetilde{Q} = \prod_{i \in R} \frac{(N_i - \nu_i)^{N_i-\nu_i}}{N_i^{N_i}}  \prod_{j \in P} \frac{(N_j + \upsilon_j)^{N_j+\upsilon_j}}{N_j^{N_j}}
	\end{equation}
	where $R$ is the set of reactant species and $P$ is the set of product species.  Lastly, defining $\Delta G^0$ as the standard state change in Gibbs free energy, we arrive at the following expression for the change in Gibbs free energy as a function of the instantaneous vector of species copy number $\bold{N}$:
	\begin{equation}
		\label{eq10}
		\Delta G(\bold{N}) = \Delta G^0 - \sigma k_B T  \log{\Theta} - \sigma k_B T + k_B T \log{\widetilde{Q}}.
	\end{equation}
	Although this equation might appear unusual, upon making some approximations leveraging the relative size of the stoichiometric coefficients and the species copy numbers it can be shown to reduce to the familiar textbook expression
	\begin{equation}
		\label{eq11}
		\Delta{G}(\bold{C}) = \Delta G^0 + k_B T \log{Q}
	\end{equation}
	where 
	\begin{equation}
		Q = \prod_{i \in R} C_i^{-\nu_i} \prod_{j \in P} C_j^{\upsilon_j}
		\label{eq11a}
	\end{equation}
	is the reaction quotient of the species concentrations $C_i$, and $\bold{C}$ is the vector of these concentrations.  In the context of a compartment-based reaction diffusion scheme, the species copy numbers used in Equation \ref{eq10} are specific to the compartment in which the reaction was drawn.  We prefer to use Equation \ref{eq10} in simulation because small deviations from this nearly exact expression can actually lead to significant systematic bias of the change in Gibbs free energy resulting from certain reactions.  This is especially true for reactions that are very frequent, such as the diffusion reaction between adjacent compartments.   $\Delta G$ for diffusion reactions, in which a molecule jumps from a compartment where its concentration is $N_{i,A}$ to a compartment where its concentration is $N_{i,B}$, is calculated using an expression similar to Equation \ref{eq10}: 
	\begin{equation}
		\Delta G = k_B T \log \frac{(N_{i,A}-1)^{(N_{i,A}-1)}}{N_{i,A}^{N_{i,A}}} \frac{(N_{i,B}+1)^{(N_{i,B}+1)}}{N_{i,B}^{N_{i,B}}}.
		\label{eq11b}
	\end{equation}
	Similarly this expression reduces to the familiar formula 
	\begin{equation}
		\Delta G = k_B T \log \frac{N_{i,B}}{N_{i,A}}
		\label{eq11c}
	\end{equation}
	upon leveraging the sizes of $N_{i,A}$ and $N_{i,B}$ relative to 1.  The derivation of the above results can be found in an accompanying paper \cite{2019arXiv190110520F}.
	
	For some of the reactions in the CT model, Equation \ref{eq11} can be directly applied and straightforwardly cast into the form of Equation \ref{eq10} to give more exact results.  However other reactions including (de)polymerization, myosin filament walking, myosin filament (un)binding, and cross-linker (un)binding, require some additional treatment. 
	
	For (de)polymerization reactions, care should be taken in defining the concentrations of the reactants and products since those molecules include heteropolymers; we would like to avoid requiring that a specific sequence of distinct subunits constitutes a unique chemical species.  This is possible to do because the polymerization of actin subunits is independent of the chemical identity of the subunit at the tips of the polymer, and is therefore independent of the chemical identity of any of the polymerized subunits \cite{brooks2009nonequilibrium}.  Then, polymerization potentially only depends on the polymer length, such that a general reaction for reversible polymerization can be written as 
	\begin{equation}
		\label{eq14}
		(m)_n + m \longleftrightarrow (m)_{n+1}
	\end{equation}
	where $(m)_n$ is a polymer with degree of polymerization $n$ and $m$ is a subunit.  For this reaction Equation \ref{eq11} reads
	\begin{equation}
		\label{eq15}
		\Delta G = \Delta G^0 + kT \log \frac{C_{(m)_{n+1}}}{C_{(m)_n}C_{m}}.
	\end{equation}  
	Following \cite{duda2009thermodynamics}, we make the simplifying assumption that a polymer's chemical reactivity does not depend on its degree of polymerization, provided that the polymer is sufficiently long and cooperative effects do not apply.  With this, Equation \ref{eq15} reduces to 
	\begin{equation}
		\label{eq16}
		\Delta G = \Delta G^0 + kT \log \frac{1}{C_{m}}.
	\end{equation}  
	This assumption agrees more or less with intuition: the polymerizing subunit does not really ``see" the degree of polymerization of the polymer, which is reflected by constant rates of polymerization for polymers of varying lengths. 
	
	Cross-linking proteins are incorporated in MEDYAN as diffusing species that bind and mechanosensitively unbind to pairs of actin filaments, mechanically coupling them.  We treat the change in free energy of the (un)binding reactions analogously to the (de)polymerization reactions.  Ignoring the chemical identities of the heteropolymer subunits, a general cross-linking protein (un)binding reaction can be written as 
	\begin{equation}
		(m)_i + (m)_j + L \longleftrightarrow (m)_i L (m)_j
		\label{eq17}
	\end{equation}
	where $L$ is a cross-linking protein and $(m)_i$ is a polymer with degree of polymerization $i$.  According to Equation \ref{eq11}, the change in Gibbs free energy upon this reaction occurring to the right is
	\begin{equation}
		\Delta G = \Delta G^0 + k T \log \frac{C_{(m)_iL(m)_j}}{C_{(m)_i}C_{(m)_j}C_{L}}.
		\label{eq18}
	\end{equation}
	Similarly to our assumption that the chemical reactivity of polymers is independent of degree of polymerization for sufficiently long polymers, we assume here that the binding affinity of cross-linking proteins is independent of the degree of polymerization and of the number of cross-linking proteins already bound to the pair of filaments, allowing us to simplify Equation \ref{eq18} to 
	\begin{equation}
		\Delta G = \Delta G^0 + k T \log \frac{1}{C_{L}}.
		\label{eq19}
	\end{equation}
	
	The kinetics of myosin filament walking in MEDYAN are based on the Parallel Cluster Model (PCM) \cite{erdmann2013stochastic}.  Taking the rate constants of individual myosin head (un)binding reactions as inputs, and accounting for the statistical distribution of the number of bound heads as well as dependence on the force exerted on the filament, the results of the PCM allow us to write kinetic parameters describing the entire myosin filament, including filament (un)binding and walking rates.  In the MEDYAN implementation, each step of a myosin filament can represent several steps of the constituent myosin heads, where each head step represents the completion of a single myosin head cross-bridge cycle \cite{erdmann2016sensitivity, howard2001mechanics}.  The head steps have a fixed length, $d_\text{step}$, set by experimental measurements of the myosin isoform of interest.  The length of the filament steps, $d_\text{total}$, is determined from the following MEDYAN parameters: the equilibrium length $L_\text{cyl}$ of the cylinders that comprise the coarse-grained representation of actin filaments, and the number of binding sites per cylinder $N_\text{bs}$, giving $d_\text{total} = L_\text{cyl}/N_\text{bs}$.  The binding sites represent the discrete locations on the cylinders which can be occupied by cross-linkers or myosin filaments.  To account for the discrepancy between $d_\text{step}$ and $d_\text{total}$, we multiply the filament walking rate by the ratio
	\begin{equation}
		s = \frac{d_\text{step}}{d_\text{total}}.
		\label{eq20}
	\end{equation}
	When a filament step occurs in MEDYAN, it thus represents the completion of $s^{-1}$ myosin head cross-bridge cycles, each of which has the effect of converting one solvated ATP molecule into solvated Pi and ADP.  The Gibbs free energy the filament walking reaction is then
	\begin{equation}
		\Delta G = s^{-1}\left( \Delta G^0 + k_B T \log \frac{C_{ADP}C_{Pi}}{C_{ATP}}   \right).
		\label{eq21}
	\end{equation}
	where $\Delta G^0$ refers to the standard change in Gibbs free energy for hydrolysis of ATP.  This expression for $\Delta G$ could be further multiplied by a parameter $\zeta$ representing the coupling of the ATP hydrolysis cycle to the forward step of the myosin head; here we follow the assumption of tight coupling, i.e. $\zeta = 1$  \cite{sakamoto2008direct, howard2001mechanics}.  
	
	The mechanochemical updating of force-sensitive reaction rates $k_\pm$ alters the equilibrium constant via $K_\text{eq} = k_-/k_+$, and therefore the change in free energy $\Delta G$ corresponding to that reaction via $\Delta G^0 = k_B T \log K_\text{eq}$.  For a reaction whose rates have been updated due to some applied force $F$, it can be shown that the new value of $\Delta G^0$ is approximately given by
	\begin{equation}
		\label{eq13}
		\Delta G^0(F) = \Delta G^0|_{F=0} + g(F)
	\end{equation}
	where $G^0|_{F=0}$ is the original (i.e. zero-force) value of $\Delta G^0$ and $g(F)$ is equal to the increase mechanical energy due to the applied force \cite{keller2000mechanochemistry, howard2001mechanics}.  In this modeling, the extra energy $g(F)$ is counted as a part of $\Delta G_\text{mech, dissipated}$, not $\Delta G_\text{chem, dissipated}$.  So when, for example, a cross-linking protein unbinds under tension $F$, the zero-force value $\Delta G^0|_{F=0}$ is used when computing the change in free energy for that reaction, and when mechanical equilibration next occurs, the released stretching energy $g(F)$ is included in the calculation of $\Delta G_\text{mech, dissipated}$. 
	
	\subsection{Parameterization}
	Parameterization of the CT model for the purpose of tracking free energy changes during simulation trajectories consists of choosing values of the rate constants (kinetic parameters) and of $\Delta G^0$ (thermodynamic parameters) for all reactions in the model.  Wherever possible, values from the literature are used.  Experimental measurements have determined rate constants for every reaction, however for some reactions the value of $\Delta G^0$ hasn't been reliably measured, to the best of our knowledge.  Below, we describe a technique to solve for these unknown values.
	
\begin{figure}[H]
		\centering
		\includegraphics[width=13 cm]{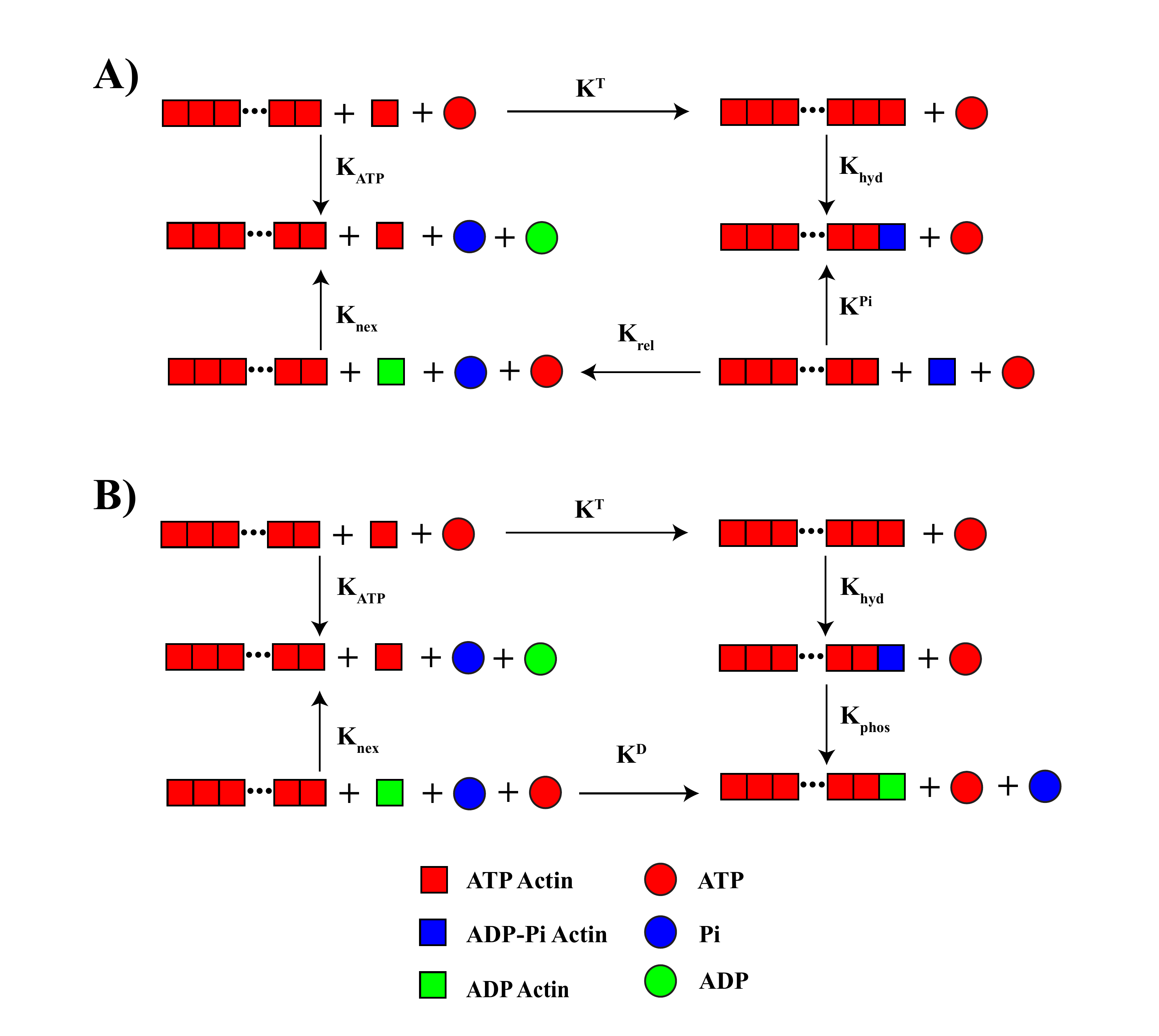}
		\caption{Diagrams representing sequences of reactions, with the involved species drawn between each reaction, resulting in independent relations between the equilibrium constants.  The meaning of these equilibrium constants is provided in the main text.  Loops are assumed to proceed in the clockwise direction, and arrows that point opposite to this direction indicate that that reaction is occurring ``backwards", i.e. from products to reactants. Polymers are shown as connected chains of subunits, while the ``+" sign indicates different solvated species.}
		\label{EnergyCircles}
	\end{figure}

	For reversible reactions, where the forward and reverse rate constants $k_+$ and $k_-$ are known, such as (de)polymerization and (un)binding of cross-linkers, $\Delta G^0$ can be found from 
	\begin{equation}
		\Delta G^0 = k_B T \log K_\text{eq} = k_B T \log \frac{k_-}{k_+}
		\label{eqb1}
	\end{equation}
	Literature values for the equilibrium constants or of $\Delta G^0$ are used for irreversible reactions, for which $k_-$ is often too small to determine from direct measurement.  Irreversible reactions in this system include all reactions except for (de)polymerization reactions.  For reactions for which literature values of $\Delta G^0$ are unavailable, it is possible to solve for $\Delta G^0$ values based on a self-consistency condition \cite{dufort1996profilin, yarmola2008effect}: the sum of the $\Delta G^0$ values around a closed loop of reactions in which the number of molecules has not experienced a net change must be zero since the free energy is a state function (equivalently, by Equation \ref{eqb1}, the product of equilibrium constants around any such loop must be equal to one).  Writing several such closed loops of reactions leads to a system of equations that can be solved for the unknown variables.  Not all possible loops result in independent equations, but we were able to determine the values of two unknown parameters using the loops illustrated in SI Figure \ref{EnergyCircles}.  
	
	The loops in SI Figure \ref{EnergyCircles} imply the following independent system of equations:
	\begin{align}
		K^T K_\text{hyd} K_\text{rel} K_\text{nex} = K^{Pi} K_\text{ATP}  \label{eqb2} \\
		K^T K_\text{hyd} K_\text{phos} K_\text{nex} = K^D K_\text{ATP} \label{eqb3}
	\end{align}
	where $K_\text{hyd}$ represents the hydrolysis of ATP by $F^T$, $K_\text{phos}$ represents the release of Pi by $F^{Pi}$, $K_\text{nex}$ represents nucleotide exchange converting $G^D$ to $G^T$, $K_\text{ATP}$ represents the hydrolysis of ATP in solution producing ADP and Pi, and $K_\text{rel}$ represents the release of phosphate by $G^{Pi}$.  
	
	Values from the literature \cite{fujiwara2007polymerization, mccullagh2014unraveling, zhang2016thermodynamic} can be used to determine 6 of the 8 variables in Equations \ref{eqb2} and \ref{eqb3}, which thus represent two equations in two unknowns: $K_\text{rel}$ and $K_\text{nex}$.  The resulting parameters are listed in Table \ref{table1}.  
	
	Note that it is possible to draw loops such as those in SI Figure \ref{EnergyCircles} that would imply that the equilibrium constants for polymerization and depolymerization of, for example $G^T$, should be the same at the plus and minus ends of the filaments.  This condition is not borne out by the experimental values of these equilibrium constants, and this discrepancy is a recognized outstanding problem \cite{fujiwara2007polymerization}.  Here, we use the literature values for these equilibrium constants and employ the reaction loop method only to determine the parameters $K_\text{rel}$ and $K_\text{nex}$.
	
	For reversible binding of myosin filaments, results from the PCM are used to describe binding and unbinding rates, and therefore $K_\text{eq}$  \cite{erdmann2013stochastic, erdmann2016sensitivity}.  The filament binding rate is given as 
	\begin{equation}
		k_\text{fil, bind} = N_t k_\text{head, bind},
		\label{eqbpcm1}
	\end{equation}
	and the unbinding rate is 
	\begin{equation}
		k_\text{fil,unbind} = N_t k_\text{head, bind} \bigg[ \bigg( \frac{k_\text{head, bind}+k_\text{head, unbind}}{k_\text{head, unbind}} \bigg)^{N_t} -1 \bigg]^{-1},
		\label{eqbpcm2}
	\end{equation}
	where $k_\text{head, bind}$ and $k_\text{head, unbind}$ describe the binding kinetics of a single myosin head, and $N_t$ is the number of heads in the filament \cite{erdmann2013stochastic}.  The resulting expression for $\Delta G^0$ is
	\begin{equation}
		\Delta G^0 = k T \frac{k_\text{fil, unbind}}{k_\text{fil, bind}} = - kT \log \bigg[ \bigg( \frac{k_\text{head, bind}+k_\text{head, unbind}}{k_\text{head, unbind}} \bigg)^{N_t} -1 \bigg].
		\label{eqbpcm3}
	\end{equation}

	We assume that we have chemostatted concentrations of ATP, ADP, and Pi, which we account for implicitly via the effect of these concentrations on the kinetic and thermodynamic parameters of certain reactions.  Thus these species are not explicitly tracked.  The concentrations of these species affect the change in Gibbs free energy associated with the following reactions:
	\begin{itemize} 
		\item myosin filament walking
		\item nucleotide exchange ($G^D \rightarrow G^T$)
		\item phosphate release by F and G-actin ($F^{Pi} \rightarrow F^D$, $G^{Pi} \rightarrow G^D$)
	\end{itemize}
	The effect of the concentrations of ATP, ADP, and Pi for the these reactions is to simply change the reaction quotient $Q$ which changes $\Delta G$ via
	$\Delta{G} = \Delta G^0 + k_B T \log{Q}$.  For instance the nucleotide exchange reaction can be written explicitly as 
	\begin{equation}
		ATP + G^D \rightarrow ADP + G^T
		\label{b4}
	\end{equation} 
	and the change in Gibbs free energy is 
	\begin{equation}
		\Delta G = \Delta G^0 + k_B T \log \left( \frac{C_{ADP} C_{G^T}}{C_{ATP}C_{G^D}} \right).
		\label{eqb5}
	\end{equation}
	To treat the concentrations of ATP and ADP implicitly, we rewrite the reaction as 
	\begin{equation}
		G^D \rightarrow G^T
		\label{eqb6}
	\end{equation} 
	for which the change in Gibbs free energy is 
	\begin{equation}
		\Delta G = \Delta {G^0}'  + k_B T \log \left( \frac{ C_{G^T}}{C_{G^D}} \right),
		\label{eqb7}
	\end{equation}
	where 
	\begin{equation}
		{\Delta G^0}' = \Delta G^0 +   k_B T \log \left( \frac{ C_{ADP}}{C_{ATP}} \right).
		\label{eqb8}
	\end{equation}
	A similar approach is taken for the other reactions mentioned above.

	The concentration of just ATP (since neither ADP or Pi appear implicitly as reactants in any of the reactions of the CT model) affects the kinetics of the following reactions:
	\begin{itemize} 
		\item myosin filament walking
		\item nucleotide exchange
	\end{itemize}
	To understand the effect of $C_{ATP}$ on the myosin filament walking rate, we employ the five state cross-bridge model of a single myosin head described in \cite{erdmann2016sensitivity}.  In that model unbinding of a head from the filament substrate occurs via two pathways: a slip path, with rate $k_{35}$, and a catch path, with an effective rate $k_{345}$.  The catch path is a two-step reaction: the release of ADP with rate $k_{34}$, followed by the unbinding of the filament head and binding of ATP with rate $k_{45} = k_T C_{ATP}$.  The effective rate constant of the approximate one-step representation of this reaction is 
	\begin{equation}
		k_{345} = \frac{k_{34} k_T C_{ATP}}{k_{34} + k_T C_{ATP}}.
		\label{eqb9}
	\end{equation}
	Unless $C_{ATP}$ is very low, this reaction rate is limited by $k_{34}$.  Because the head can unbind by the catch or slip pathway, the rate for unbinding is 
	\begin{equation}
		k_\text{head, unbind} = k_{35} + \frac{k_{34} k_T C_{ATP}}{k_{34} + k_T C_{ATP}}
		\label{eqb10}
	\end{equation}
	The slip path should also be dependent on ATP concentration, since in state 5 the head is ATP-bound, however we follow the authors of \cite{erdmann2016sensitivity} in neglecting this dependence since the slip path only becomes active under large load.
	
	The rate of the nucleotide exchange reaction also depends on $C_{ATP}$ and also occurs in two steps.  The full reaction, with ATP and ADP explicitly included, is 
	\begin{equation}
		G^D + ATP \rightarrow G^* + ATP + ADP \rightarrow G^T + ADP
		\label{eqb11}
	\end{equation}
	where $G^*$ represents actin with no bound nucleotide.  Following \cite{brooks2009nonequilibrium}, this reaction is approximated as a one-step irreversible reaction with ATP and ADP included explicitly, Equation \ref{eqb6}. We can write the rate $k_\text{nex}$ of this approximate reaction as:
	\begin{equation}
		k_\text{nex} = \frac{k_{D\rightarrow*}k_{*\rightarrow T}C_{ATP}}{k_{D\rightarrow*}+k_{*\rightarrow T}C_{ATP}} \approx k_{D\rightarrow*}
		\label{eqb12}
	\end{equation}
	where $k_{D\rightarrow*}$ is the dissociation rate of ADP, $k_{*\rightarrow T}$ is the second order rate constant of ATP association, and the approximation holds except at low concentrations of ATP when ADP dissociation is no longer the rate-limiting step.  The values of $k_{*\rightarrow T}$ and $k_{D\rightarrow*}$ are presented and discussed in \cite{nowak1988kinetics, kinosian1993nucleotide, selden1999impact}.

	\begin{table}[H]
		\begin{center}
			
			\begin{tabular}{ | p{5cm} | l | l | l |}
				\hline 
				\textbf{Reaction} & \textbf{Rate Constant} & $\Delta G^0$ ($k_BT$ ) & \textbf{Reference}  \\ \hline
				$G^T$ poly at plus end & 11.6 $(\mu M s)^{-1}$ &  -2.12 $^\bold{a}$ & \cite{brooks2009nonequilibrium} \\
				$G^T$ depoly at plus end & 1.4 $ s^{-1}$ &  2.12 $^\bold{a}$  & \cite{brooks2009nonequilibrium} \\
				$G^T$ poly at minus end & 1.3 $(\mu M s)^{-1}$ &  -0.51 $^\bold{a}$  & \cite{brooks2009nonequilibrium}  \\
				$G^T$ depoly at minus end & 0.8 $ s^{-1}$ &  0.51 $^\bold{a}$ & \cite{brooks2009nonequilibrium}  \\
				
				$G^{Pi}$ poly at plus end & 3.4 $(\mu M s)^{-1}$ &  -2.81 $^\bold{a}$ & \cite{fujiwara2007polymerization}  \\
				$G^{Pi}$ depoly at plus end & 0.2 $ s^{-1}$ &  2.81 $^\bold{a}$ & \cite{fujiwara2007polymerization}  \\
				$G^{Pi}$ poly at minus end & 0.11 $(\mu M s)^{-1}$ &  -1.75 $^\bold{a}$ & \cite{fujiwara2007polymerization}  \\
				$G^{Pi}$ depoly at minus end & 0.02 $ s^{-1}$ &  1.75 $^\bold{a}$ & \cite{fujiwara2007polymerization}  \\
				
				$G^D$ poly at plus end & 2.9 $(\mu M s)^{-1}$ &  0.59 $^\bold{a}$ & \cite{brooks2009nonequilibrium} \\
				$G^D$ depoly at plus end & 5.4 $ s^{-1}$ &  -0.59 $^\bold{a}$ & \cite{brooks2009nonequilibrium} \\
				$G^D$ poly at minus end & 0.09 $(\mu M s)^{-1}$ &  1.03 $^\bold{a}$ & \cite{brooks2009nonequilibrium} \\
				$G^D$ depoly at minus end & 0.25  $ s^{-1}$ &  -1.03  $^\bold{a}$ & \cite{brooks2009nonequilibrium} \\
				
				Pi release by F-actin & 0.002 $ s^{-1}$ &  -7.31 $^\bold{b, e}$ & \cite{brooks2009nonequilibrium, fujiwara2007polymerization} \\
				ATP hydrolysis by F-actin & 0.3 $ s^{-1}$ &  -10.0 & \cite{brooks2009nonequilibrium, mccullagh2014unraveling} \\
				Pi release by G-actin & 5  $ s^{-1}$ &  -10.77 $^\bold{c, e}$ & -  \\
				Nucleotide exchange & 0.01  $ s^{-1}$ $^\bold{e}$ &  -6.76 $^\bold{c, e}$ & \cite{brooks2009nonequilibrium} \\
				Cross-linker binding & 0.7 $(\mu M s)^{-1}$ & -0.85 $^\bold{a}$ & \cite{popov2016medyan} \\
				Cross-linker unbinding & 0.3 $s^{-1}$ & 0.85 $^\bold{a}$ & \cite{popov2016medyan} \\
				Myosin head binding & 0.2 $s^{-1}$ $^\bold{d}$ & -  $^\bold{d}$ & \cite{popov2016medyan} \\
				Myosin head unbinding & 1.708 $s^{-1}$ $^\bold{d}$ & - $^\bold{d}$ & \cite{popov2016medyan} \\
				Myosin filament walking & - $^\bold{f}$ & -14.5 $^\bold{e}$ & \cite{milo2015cell}\\
				\hline
				
			\end{tabular}
			\caption{Kinetics and thermodynamic parameters describing reactions in the CT model as well as cross-linker and myosin filament (un)binding and myosin filament walking.}\label{table1}

		\end{center}
	\end{table}
	
	$^\bold{a}$ - Values of $\Delta G^0$ determined via Equation \ref{eqb1}.
	
	$^\bold{b}$ - $\Delta G^0$ determined from $K_\text{eq}$ given in \cite{fujiwara2007polymerization}.
	
	$^\bold{c}$ - $\Delta G^0$ determined using constraints as described above.
	
	$^\bold{d}$ - Parameters describing the myosin filament obtained via Equations \ref{eqbpcm1}, \ref{eqbpcm2}, \ref{eqbpcm3}. 
	
	$^\bold{e}$ - Depends implicitly on $C_{ATP}$, $C_{ADP}$, and $C_{Pi}$; given value applies to standard state.  
	
	$^\bold{f}$ - Calculated in simulation using results of PCM, see \cite{popov2016medyan}.

	In all the studies in this paper, implicit nucleotide concentrations are taken to be $C_{ATP}$ = 8 $m M$, $C_{ADP}$ = 7 $\mu M$, and $C_{Pi}$ = 1 $m M$, corresponding roughly to the amounts found in human muscle after exercise \cite{milo2015cell}.  Other parameters of the system, including mechanical constants, diffusion rates, screening lengths, and boundary cutoffs have been set to the same values given and discussed in \cite{popov2016medyan}.  Cylinder equilibrium lengths $L_\text{cyl}$ are chosen as 27 $nm$ with 4 binding sites per cylinder for myosin filaments and 1 binding site per cylinder for cross-linkers, giving approximately physiological values for stepping distances of myosin motor filaments and spacing along actin filaments of bound $\alpha$-actinin.  We note that the form of the mechanochemical models has been changed from those used in \cite{popov2016medyan}; the modeling used here is current as of MEDYAN v3.2, and we refer readers to the documentation at http://www.medyan.org/ for further details.

	\subsection{Mean-Field Model of Treadmilling Dissipation}
	
	To validate the methods for quantifying dissipation using MEDYAN against a simpler representation of actin filament treadmilling, we developed a mean-field description of the dissipation resulting from chemical reactions in the CT model.  Mean-field models of the trajectory of the vector of concentrations of species, $\bold{C}(t)$, have been formulated previously as an 11-dimensional system of ordinary differential equations (ODEs) (\cite{brooks2009nonequilibrium}), and in the CT model as a 5-dimensional system of ODEs  (\cite{floyd2017low}).  These models describe the polymerization of a concentration $N_\text{fil}$ of actin filaments in a pool of actin subunits of total concentration $M$.  The subunit species tracked by these models are distinguished by their polymerization state and by the hydrolysis states of the nucleotide to which they are bound. The meaning of ``mean-field" in this context is the assumption that reacting species are well-mixed over the entire system volume, therefore obeying mass-action kinetics and deterministic dynamics which can be represented by ODEs.  For an instantaneous value of $\bold{C}$, we define a function $D_\Lambda(\bold{C})$ representing the instantaneous rate of dissipation due to a set of reactions $\Lambda$.  Thus a solution of a mean-field $\bold{C}(t) = \bold{N}(t)/\Theta$, allows us to construct the trajectory of the dissipation rate, $D_\Lambda(\bold{C}(t))$.  This function cannot capture the dissipation due to the activity of myosin filaments or cross-linkers, or from relaxation of mechanical stress, because these aspects are not included in the these mean-field models describing filament treadmilling.  The benefit of such a mean-field model is that one can perform systematic variation of parameters with limited computational demands, and we use it here to study the effect of the parameters $N_\text{fil}$ and $M$ on the dissipation due to filament treadmilling.
	
	The function $D_\Lambda(\bold{C}(t))$ can be written as a sum over the reactions $\lambda \in \Lambda$ of the instantaneous rate of change of the solution's Gibbs free energy due to that reaction:
	
	\begin{equation}
		D_\Lambda(\bold{C}(t)) = \sum_{\lambda \in \Lambda} \Delta G_\lambda(\bold{C}(t)) r_\lambda(\bold{C}(t)).
		\label{eq22}
	\end{equation}
	The expressions for $\Delta G_\lambda(\bold{C}(t))$ for different reactions are described above.  The instantaneous rate of reaction $\lambda$, $r_\lambda(\bold{C}(t))$, is written as usual for mass-action kinetics as
	\begin{equation}
		r_\lambda(\bold{C}(t)) =\Theta k_\lambda \prod_{i \in R} C_i^{\nu_i}
		\label{eq23}
	\end{equation}
	where $R$ is the set of reactant species for reaction $\lambda$, and where we have included the conversion factor $\Theta$ to convert $r_\lambda(\bold{C}(t))$ to units of $s^{-1}$.
	
	To facilitate the study of dissipation due to filament treadmilling, we define a set of reactions $\Lambda$ in the CT model which constitutes the dominant cycle a subunit undergoes in the treadmilling process: a) polymerization of $G^T$ to the plus end, b) hydrolysis of ATP by $F^T$, c) release of Pi by $F^{Pi}$, d) depolymerization of $G^D$ from the minus end, and e) nucleotide exchange converting $G^D$ to $G^T$.  We refer to this set of 5 reactions as the main treadmilling pathway (MTP).  Alternative sequences of reactions whose net effect is similarly the conversion of one molecule of ATP to ADP and Pi are considered as of secondary importance and not included in this analysis.

\begin{center}
		\begin{figure}[H]
			\centering
			\includegraphics[width= 15 cm]{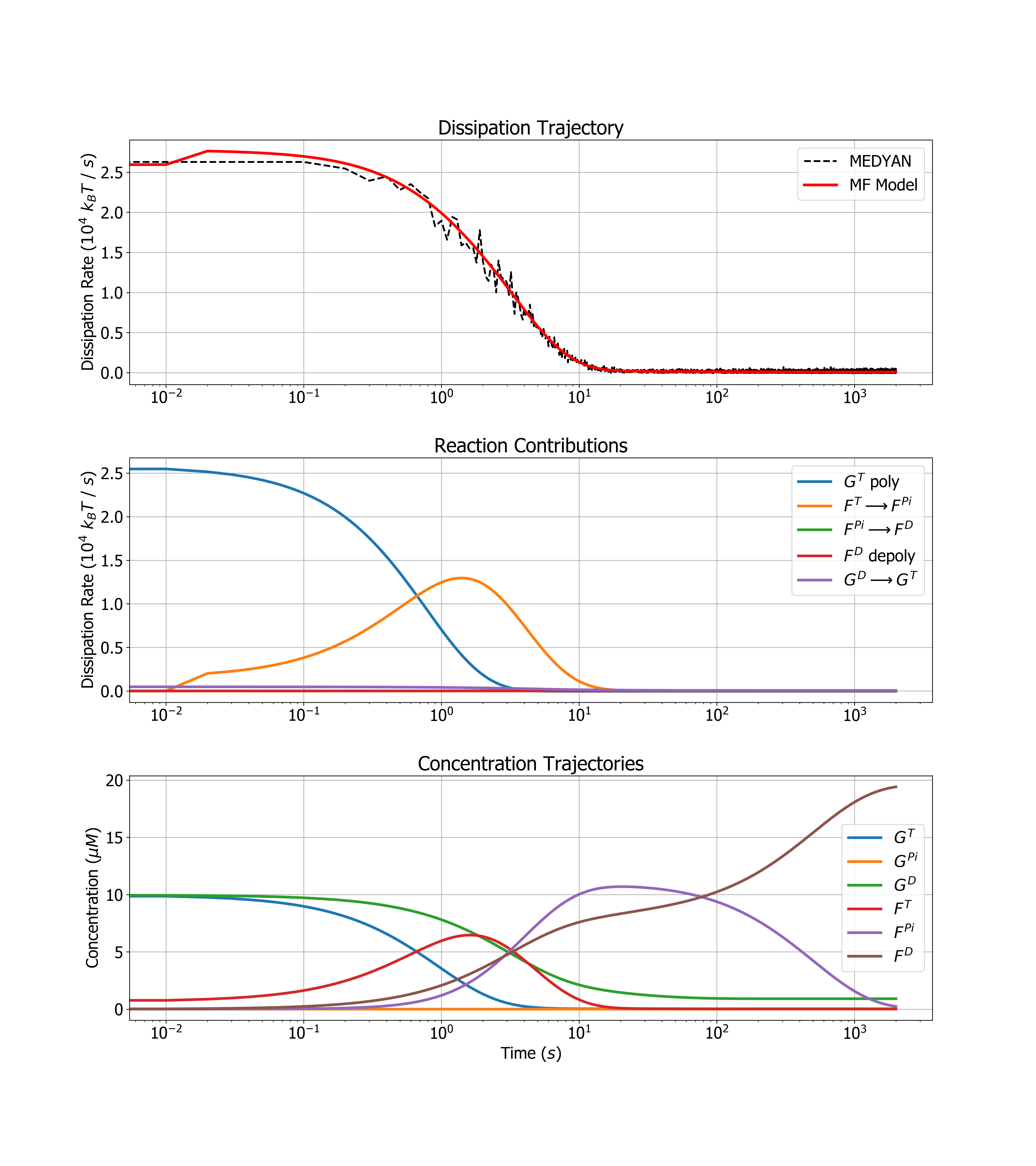}
			\caption{Results of the mean-field modeling of MTP dissipation.  \textit{Top}:  Comparison of $D_\text{MTP}(\bold{C}(t))$ calculated using the mean-field model (with time increments of 0.01 s), with $\Delta G_\text{chem}$ rates measured during MEDYAN simulation (with time increments of 0.1 s). \textit{Middle}:  Plot of the contributions of each reaction in the MTP to the total dissipation.  The items in the legend represent reactions in the MTP, which are described in the main text.
				\textit{Bottom}:  Plot of the trajectories of the  reacting species concentrations.  The notation for each species is described in the main text. }
			\label{combined}
		\end{figure}
	\end{center}

	This mean-field description of entropy production can be compared to the description of entropy production rates for chemically reactive systems that emerged from the Brussels school of thermodynamics.  \cite{nicolis1977,kondepudi2014modern}.  The results of that school include the minimum entropy-production principle applicable in the linear regime \cite{prigogine1967introduction}, and the general evolution criterion applicable even in the the non-linear regime \cite{glansdorff1954proprietes}.  Their formalism typically considers the total entropy production rate as a volume integral over the local entropy production rate density, which is itself written as a sum over fluxes multiplied by their corresponding thermodynamics forces defined at each point of the system.  This sum is decomposed into terms representing diffusion and terms representing chemical reactions, and the terms representing the chemical reactions are written such that the fluxes reflect the net reaction rate at that point in the system.  In contrast, our mean-field description neglects concentration gradients and resulting diffusion fluxes, as we assume a homogeneous distribution of the chemical species.  Equation \ref{eq22} then represents only the chemical contribution to the entropy production, as a sum over the rates and affinities of the reactions in the system, implicitly integrating over the homogeneous system volume.  We treat the forward and reverse direction for some chemical reaction as separate terms in Equation \ref{eq22}, so these rates cannot be considered fluxes which would include the reverse rate as well.  This allows for more general sets of reactions which might include effectively irreversible processes for which the reverse rate is negligible.  However the set of reactions $\Lambda$ could be chosen to include a reverse reaction for each forward reaction, with the result that these pairs of terms represent fluxes along the reversible reaction pathway.  The parsimonious reaction set MTP is chosen not to fully describe the rate of entropy production in the system, but to allow easy visualization of the main contributions to the entropy production.  We do not pursue the connection of our treatment to the formalism of the Brussels school further here, but we lastly note that our results are compatible with their minimum entropy-production principle, as shown in SI Figure \ref{combined}.

	We first verified that the mean-field model of MTP dissipation agreed with results from MEDYAN simulations, to illustrate consistency between these approaches.  In SI Figure \ref{combined} we display the close match between the trajectory of MTP dissipation over a 2000 $s$ run from these two approaches. Note that, to allow direct comparison, only the changes of Gibbs free energy resulting from reactions in the MTP set are visualized for both approaches here, i.e. the contribution from diffusion and other non-MTP reactions in the MEDYAN simulation are not visualized.  We also turned off in MEDYAN the force-sensitive decrease in polymerization rate when the filament tips push against the simulation hard-wall boundaries, since this effect is not captured in the mean-field modeling \cite{peskin1993cellular}.  We used a simulation volume of 1 $\mu m^3$ and initial conditions of equal amounts (10 $\mu M$ each) of $G^T$ and $G^D$ actin in a 0.08 $\mu M$ pool of seed filaments containing $F^T$.  The dissipation rate decreases nearly monotonically, attaining a minimal steady-state value after tens of seconds.  In SI Figure \ref{combined} we also display the individual contributions to the sum in Equation \ref{eq22}.  Initial polymerization of $G^T$ to the plus end constitutes the majority of the initial dissipation.  As this polymerization process slows after about 1 second, the hydrolysis of ATP by the now relatively abundant $F^T$ becomes the dominant contribution.  As hydrolysis then slows after about 10 seconds, the total dissipation rate reaches a steady-state value of roughly 80 $k_B T/s$.  In SI Figure \ref{combined} we also plot the mean-field prediction of the trajectory of the reacting species' concentrations. 
	
\begin{center}
		\begin{figure}[H]
			\centering
			\includegraphics[width= 17 cm]{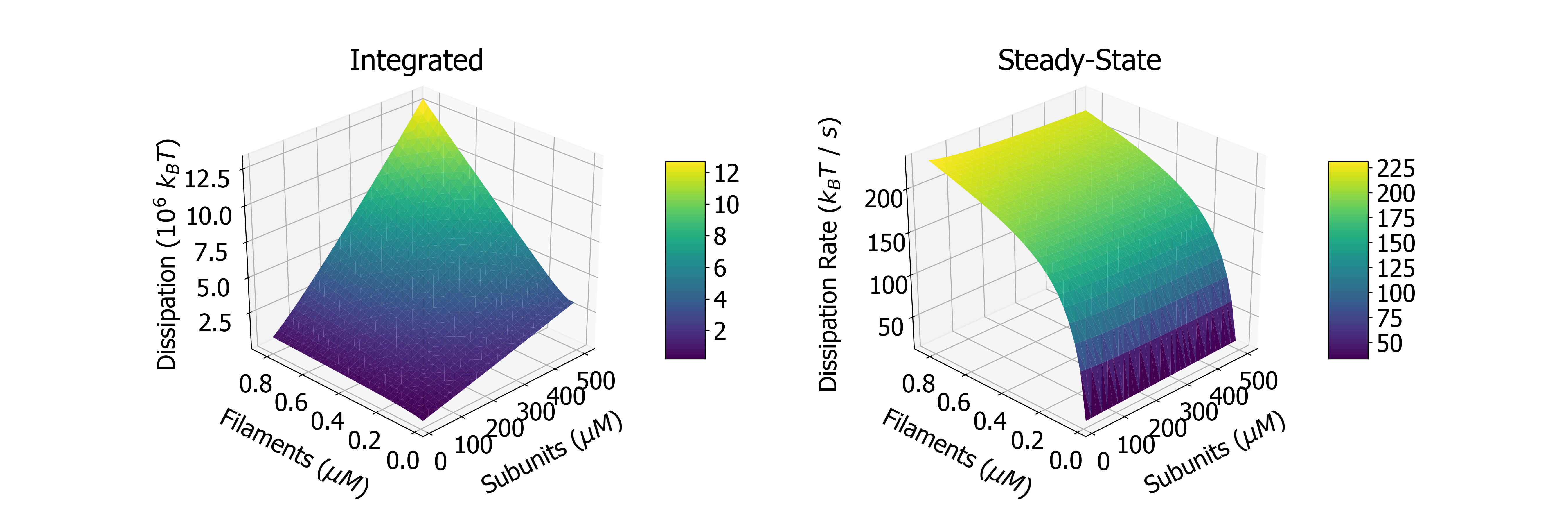}
			\caption{\textit{Left}: The total dissipation integrated over 2000 $s$ trajectories as $M$ and $N_\text{fil}$ are varied.  \textit{Right}: The steady-state dissipation rate over the same range of $M$ and $N_\text{fil}$.}
			\label{mfplots}
		\end{figure}
	\end{center}

\begin{center}
		\begin{figure}[H]
			\centering
			\includegraphics[width = 17 cm]{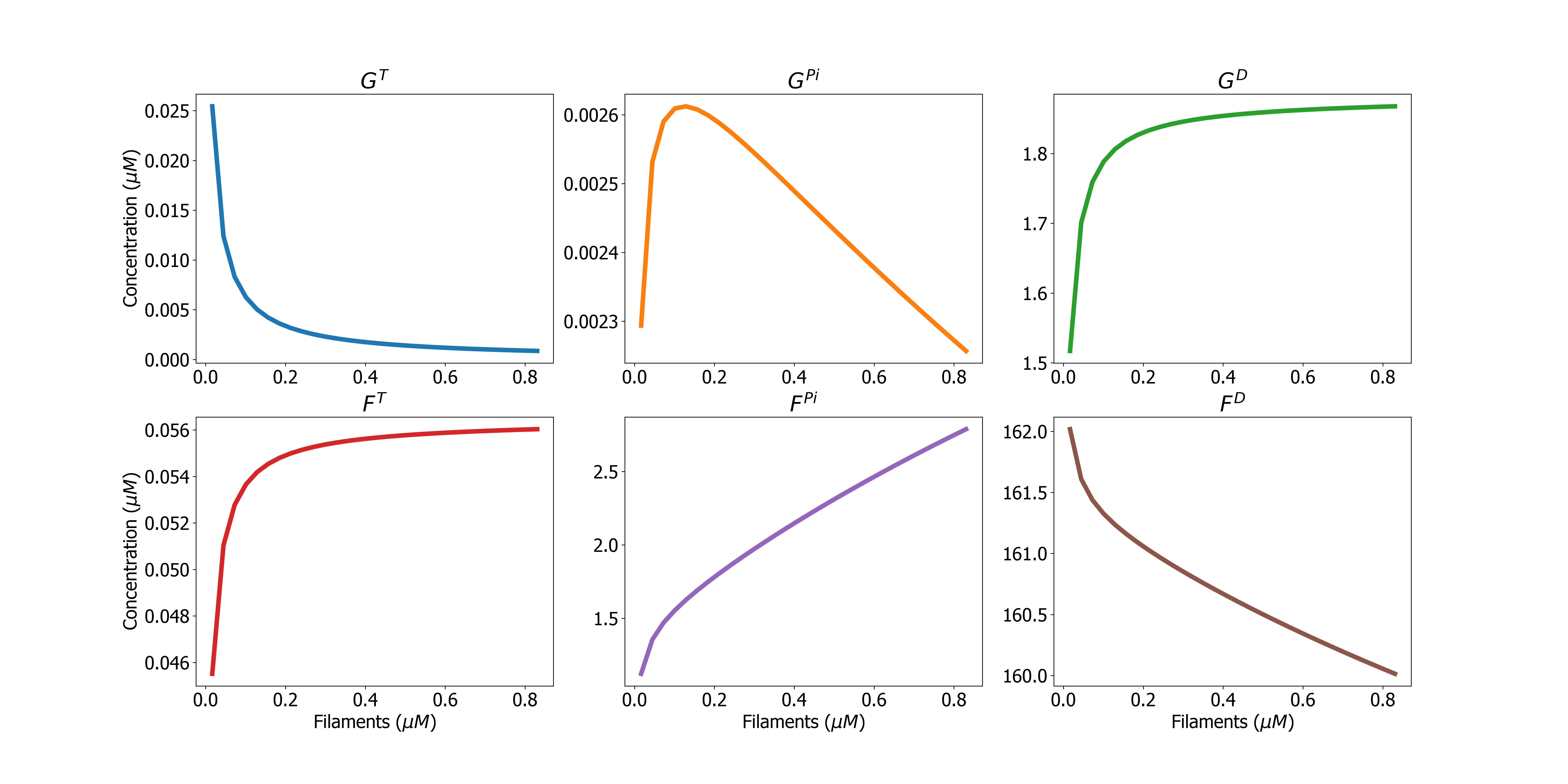}
			\caption{The concentrations at steady-state of the various actin subunit species as the concentration of filaments $N_\text{fil}$ is varied.  These curves have no dependence on initial conditions, except $F^D$ which will increase linearly with the total concentration of actin subunits $M$; any additional actin subunits in the system will accumulate in the form $F^D$ at steady-state.}
			\label{ssconcs}
		\end{figure}
	\end{center}

\begin{center}
		\begin{figure}[H]
			\centering
			\includegraphics[width= 15 cm]{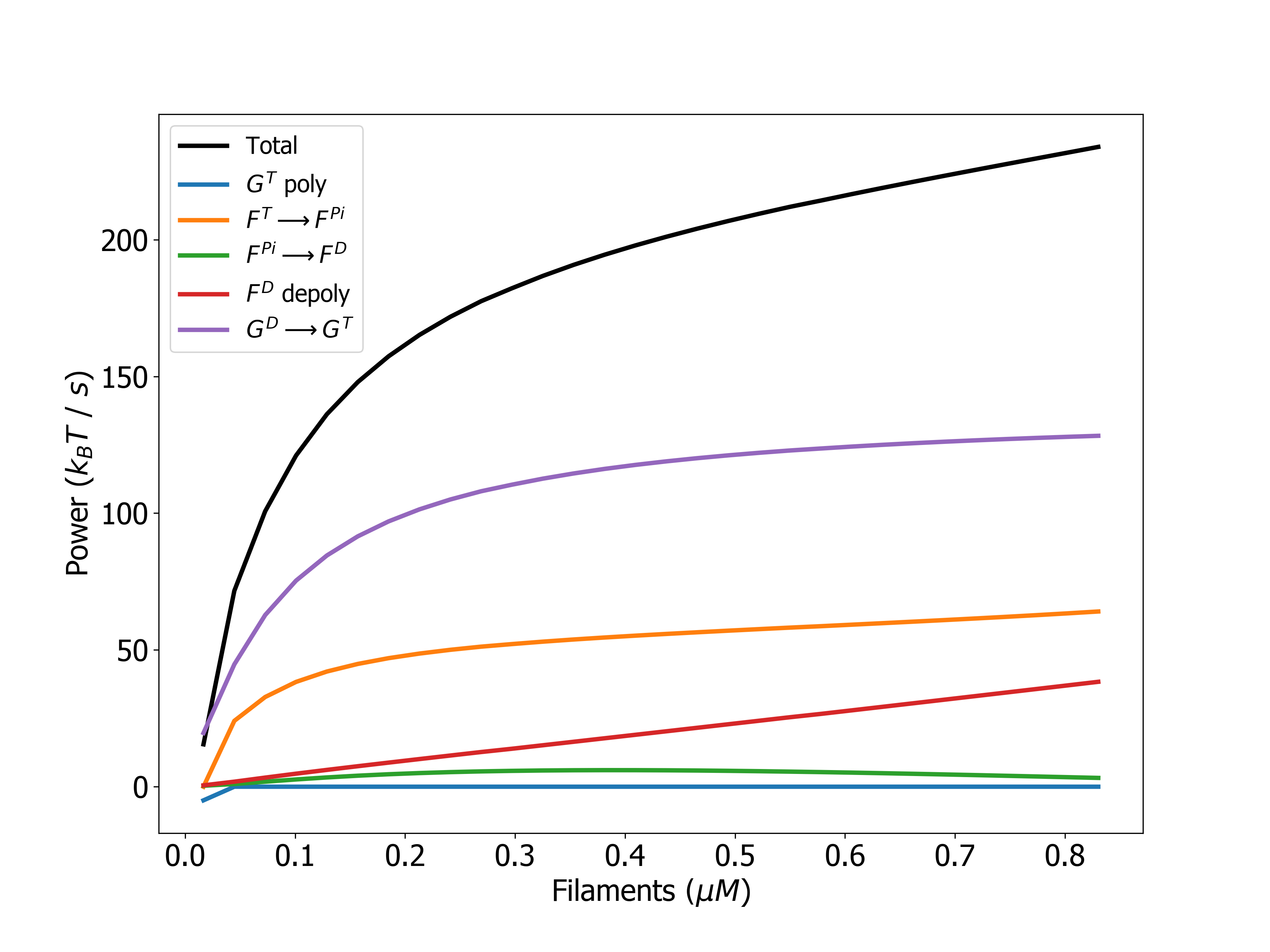}
			\caption{Contributions of each reaction in the main treadmilling pathway to the total dissipation rate at steady-state, as the concentration of filaments $N_\text{fil}$ is varied.  These curves have no dependence on the initial concentrations of the different subunit species.   }
			\label{ssconts}
		\end{figure}
	\end{center}

	We next simultaneously varied the total concentration of actin $M$ and the concentration of actin filaments $N_\text{fil}$, and determined the total dissipation integrated along each trajectory as well as the steady-state dissipation rate.  As $M$ was varied, we held the initial concentration of each species proportionally the same: 49 \% $G^T$, 49 \% $G^D$, and 2 \% $F^T$.   As shown in SI Figure \ref{mfplots}, the integrated dissipation over 2000 $s$ was observed to increase monotonically with both $M$ and $N_\text{fil}$.  Quantitatively, the integrated dissipation depends on the choice of initial proportions, however we found that the shape of the dependence on $N_\text{fil}$ and $M$ is largely independent of initial proportions (data not shown).  Total dissipation increases with $M$ simply because more actin is available to hydrolyze ATP during the approach to steady-state.  For large amounts of actin, increasing $N_\text{fil}$, the number concentration of filaments, allows increased polymerization of $G^T$, which constitutes a large contribution to total dissipation during the early stages of the trajectory.  Increasing $N_\text{fil}$ also shifts the steady-state concentration of $G^T$ downward (SI Figure \ref{ssconcs}), implying that more $G^T$ has been polymerized during the approach to steady-state.  A loose analogy can be made of a one lane road compared to a multi-lane highway during heavy traffic to describe this situation.  As $N_\text{fil}$ is increased with $M$ fixed, the steady-state dissipation rate increases concavely, as shown in SI Figure \ref{mfplots}.  The steady-state concentration of $F^T$ also increases concavely (SI Figure \ref{ssconcs}), representing higher rates of ATP hydrolysis.  The contributions of each reaction to the total dissipation rate at steady-state as $N_\text{fil}$ is varied is illustrated in SI Figure  \ref{ssconts}.  The lack of dependence of the steady-state dissipation rate on $M$ can be explained by the fact that increasing $M$ increases the steady-state concentration of only $F^D$, not of any other species \cite{floyd2017low}.  In other words, all extra actin accumulates in the form $F^D$ as $M$ is increased.  This species is essentially inert, since the depolymerization rate of $F^D$ is controlled by the concentration of filaments $N_\text{fil}$.  Thus the steady-state dissipation has no dependence on the total amount of actin.  Furthermore, it has no dependence on the initial concentrations, since it is known that the steady-state vector of concentrations does not depend on initial conditions \cite{floyd2017low}.  
	
\section*{Data accessibility}
MEDYAN is available under copyright for download at http://www.medyan.org/.  The version code and data analyzed here without mechanochemical feedback are available through the digital repository at the University of Maryland:  https://doi.org/10.13016/t8oa-9qra.  
\section*{Authors' contributions}
G.P. conceived the project. G.P. and C.J. consulted on all aspects of the project.  C.F. developed and implemented the new MEDYAN code, performed the simulations, analyzed results, and wrote the manuscript.  All authors contributed to the mean-field model, edited the manuscript, and approved it for submission.  
\section*{Competing interests}
The authors declare no competing or financial interests.  
\section*{Funding}
This work was supported by the following grants from the National Science Foundation: NSF 1632976, NSF DMR-1506969, and NSF CHE-1800418.
\section*{Acknowledgments}
The authors gratefully acknowledge A. Chandresekaran, Q. Ni, and  H. Ni for their improvements of the manuscript and helpful discussions.

\bibliographystyle{unsrt}

\end{document}